\documentstyle[prl,aps,epsf]{revtex}
\newcommand{\beq}{\begin{equation}}
\newcommand{\eeq}{\end{equation}}
\newcommand{\bqn}{\begin{eqnarray}}
\newcommand{\eqn}{\end{eqnarray}}
\newcommand{\bqns}{\begin{eqnarray*}}
\newcommand{\eqns}{\end{eqnarray*}}
\newcommand{\bary}{\begin{array}}
\newcommand{\eary}{\end{array}}
\newcommand{\non}{\nonumber}
\newcommand{\BE}{\begin{equation}}
\newcommand{\EE}{\end{equation}}
\newcommand{\BA}{\begin{eqnarray}}
\newcommand{\EA}{\end{eqnarray}}

\begin{document}

\title{Two dimensional photonic crystals}
\author{Pi-Gang Luan and Zhen Ye}
\address{Wave Phenomena Laboratory, Department of Physics,
National Central University, Chungli, Taiwan 32054}
\date{\today}
\draft \maketitle

\begin{abstract}

The topology, the symmetry involving the shape of dielectric
cylinders, and the lattice structure are among the most important
ingredients in the architecture of photonic crystals. In this
paper, we present a systematic derivation of the formulas which
are needed in computing the photonic band structures of many
commonly used two dimensional lattice structures and dielectric
cylinders with various kinds of symmetries and rotations. Further
the results are applied to arrays of hollow cross-shaped cylinders
embedded in the alumina ceramic background. A large complete
photonic band gap is found in the high frequency regime.

\end{abstract}


\section{Introduction}

Over the past ten years, the propagation of classical waves in a
periodic medium has attracted considerable interest. This includes
electromagnetic (EM) wave propagation in periodic dielectric
structures, and acoustic and elastic wave propagation in periodic
elastic composites. A new research field emerges as the wave
crystals including both photonic and sonic crystals. The photonic
or sonic crystals respectively refer to crystal-like structures
that modulate EM or acoustic wave propagation and thus lead to
dispersion bands, in analogy with the electronic energy bands in
solid state physics\cite{ssp}.

The research on photonic crystals has been particularly
intensified, after the suggestion that photonic band gaps (PBG)
could hinder spontaneous emission and block propagation of EM
waves, thus providing the possibility to manipulate the
propagation of EM waves\cite{Yab,John}. Photonic crystals offer an
unparallel opportunity to design new optical devices and hold a
great potential for many significant applications, such as
semiconductor lasers and solar cells\cite{Yab,Painter}, high
quality resonator and filters\cite{Res}, controlling photon
emission\cite{emission}, optical fibers\cite{fiber1}, guiding and
bending of EM waves with minimum losses\cite{bend,97,chow}, single
mode waveguides for light\cite{mode}, low dimensional efficient
transport of electrons and excitons by nanostructural
networks\cite{nano}, all polymer optoelectronic devices\cite{opt},
semiconductor memory cells\cite{cell}. Many methods have been
proposed for fabricating photonic crystals. These include square
spiral microfabrication achitecture for large three dimensional
band gaps\cite{Toader}, filling the voids in titania with air by
precipitation for the optical spectrum\cite{air}, using three
dimensional carbon structures\cite{carbon}, large scale synthesis
of silicon photonic crystals\cite{silicon}, fabrication of
photonic crystals for visible spectrum by holographic
lithography\cite{holo}, the electrochemical
techniques\cite{ele1,ele2}.

Indeed, the past a few years have witnessed rapid advances in both
better understanding of the exquisite properties of photonic
crystals and manipulation of EM waves by photonic crystals. A rich
body of literature on photonic crystals exists and can be found on
the internet\cite{lit}. Recently, scientists also investigated the
spine from sea mouse\cite{seamouse}. They discovered that the
spine consists of an array of regularly arranged hollow cylinders,
and this simple structure gives rise to a spectacular iridescence.
This is a remarkable example of photonic crystals by a living
oraganism.

Although three dimensional (3D) photonic crystals suggest the most
intriguing ideas for novel applications, two dimensional (2D)
structures also find several unique uses\cite{Anderson}, including
the aforementioned waveguides and communication
fibers\cite{fiber1,mode} and the 2D periodic structures in living
animals\cite{seamouse}, feedback mirror in laser
diodes\cite{diodes} and so on. In addition, fabricating 3D
periodic structures in the near infrared regime poses a
significant challenge compared to the two dimensional
situations\cite{carbon,ir,2d1,2d2}. Due to these reasons, the
study of 2D photonic crystals has been overwhelmed in the last few
years. The earliest theoretical analysis of 2D photonic band
structure was done by Plihal et al.\cite{Plihal}. The experimental
observation of photonic band structure in 2D periodic dielectric
arrays was subsequently reported by Robertson et al.\cite{WMR}.

The important issue in the fabrication of photonic crystals is to
create large, robust complete band gaps within which propagation
of EM waves is prohibited in any direction. Several methods have
been suggested for obtaining large complete band gaps in 2D
situations. For example, it has been shown that large band gaps
can be obtained by such as varying dielectric contrast ratio and
filling factors, inserting a third component into the existing
photonic crystals\cite{insertion}, reducing the structural
symmetry\cite{Anderson}, using non-circular
rods\cite{Padjen,noncircular} and subsequently by rotating the
non-circular rods\cite{rotation,chess}, rotating the lattice
structures\cite{Anderson2}, using anisotropic dielectric
materials\cite{anisotropic}, using the effects of magnetic
permeability\cite{mag}, using metallic or metallodielectric
rods\cite{met1,met2,met3}, placing rods of various shapes on
different lattice configurations such as square \cite{Plihal},
triangular\cite{anisotropic,Plihal2}, honeycomb\cite{Anderson} and
so on. Each approach may have its advantages and shortcomings. For
example, the dielectric contrast is limited by material
availability. The symmetry reduction and using non-circular rods
may reduce the degeneracy of photonic bands at high symmetry
points in the Brillouin zone, thus increasing band gaps. Although
the metallic photonic crystals can yield large band gaps, they
suffer from absorption. While the symmetry reduction method can
enhance some high order band gaps, the low order gaps are often
reduced\cite{insertion}. Therefore each method has its own
applicable situations.

Inspecting these progresses made towards better design of two
dimensional photonic crystals, we realize that the topology, the
symmetry involving the shape of dielectric cylinders, and the
lattice structure are among the most important ingredients in the
architecture of photonic crystals. The different combinations of
these factors lead to applications for various purposes. We are
therefore led to the task of deriving systematically necessary
formulas computing band structures for various configurations.
This paper is one of our attempts. In this paper, in an organized
fashion we present the analytic results for computing the band
structures of most commonly used lattice structures and dielectric
cylinders with many kinds of symmetries and rotations. With the
aim in mind that the reader can readily make use of these results,
we summarize them in tables.

This paper is organized as follows. The general theory for EM
waves in an arbitrary 2D periodic structures is presented in the
next section. Using the plane wave expansion method, the secular
equations are derived for determining the band structures for both
E-polarization mode and H-polarization mode with the electric
field and the magnetic field parallel to the longitudinal axis
respectively. In the formulation, a structure factor is
identified. Once this factor is known, the band structure can be
computed by a standard diagonalization method. In section III, the
structure factor is derived for a general configuration. Some
common crystal structures with several rod shapes are considered
in section IV. The results are presented in two tables. A
numerical example is shown in section V, followed by concluding
remarks in the last section where the extension to sonic crystals
is also discussed.

\section{Theory}

The photonic band structures can be obtained by solving Maxiwell's
equations using the plane-wave expansion method. This method can
be referred to, for example,
Refs.~\cite{Plihal,rotation,Ho,Leung}. For the sake of
convenience, we provide a brief account of this method. Consider a
periodic array of dielectric cylinders. The longitudinal axes of
the cylinders are along the $z$-axis. For dielectric materials,
Maxiwell's equations are represented in terms of the magnetic
field ${\mathbf H}$ \BE
\nabla\times\left[\frac{1}{\epsilon(\mathbf r)}\nabla\times
{\mathbf H}({\mathbf r})\right] = \frac{\omega^2}{c_0^2}{\mathbf
H}({\mathbf r}), \label{eq:ye1}\EE where $\epsilon$ is the
position-dependent dielectric constant, ${\mathbf r}$ is the
coordinates in the plane perpendicular to the rods, and $c_0$ is
the EM wave phase speed in vacuum. By Bloch theorem\cite{ssp}, the
magnetic field can be written as \BE {\mathbf H}({\mathbf r}) =
e^{i{\mathbf k}\cdot{\mathbf r}}{\mathbf H}_k({\mathbf r}),
\label{eq:ye2}\EE in which ${\mathbf H}_k({\mathbf r})$ is a
periodic function of the lattice structure, and ${\mathbf k}$ is
the Bloch wave vector within the first Brillouin zone.

Using the Fourier transformation, the magnetic field and the
dielectric function can be expanded as, \BE {\mathbf H}_{\mathbf
k}({\mathbf r}) = \sum_{\mathbf G}\sum_{j=1,2}
\hat{e}_{j}H_{j,{\mathbf k}}({\mathbf G}) e^{i{\mathbf
G}\cdot{\mathbf r}}, \label{eq:ye3}\EE and \BE \epsilon({\mathbf
r}) = \sum_{\mathbf G} \epsilon({\mathbf G}) e^{i{\mathbf
G}\cdot{\mathbf r}}, \label{eq:ye4}\EE where $\hat{e}_j$ is the
base vector for the magnetic field, ${\mathbf G}$ is a reciprocal
lattice vector. Hereafter, all quantities with a hat refer to the
unit vectors. For two dimensions, the E-polarization and
H-polarization modes are decoupled. By taking Eqs.~(\ref{eq:ye2}),
(\ref{eq:ye3}), and (\ref{eq:ye4}) into Eq.~(\ref{eq:ye1}), we
obtain two eigen-equations \BE \sum_{\mathbf G'}|{\mathbf
k+G}||{\mathbf k+G'}|\epsilon^{-1}({\mathbf G-G'})
H_{\perp,{\mathbf k}}({\mathbf G'}) =
\frac{\omega^2}{c_0^2}H_{\perp,{\mathbf k}}({\mathbf
G}),\label{eq:E}\EE for the E-polarization and \BE \sum_{\mathbf
G'}({\mathbf k+G})\cdot({\mathbf k+G'})\epsilon^{-1}({\mathbf
G-G'}) H_{\parallel,{\mathbf k}}({\mathbf G'}) =
\frac{\omega^2}{c_0^2}H_{\parallel,{\mathbf k}}({\mathbf
G}),\label{eq:H}\EE for the H-polarization. Here
$\epsilon^{-1}({\mathbf G-G'})$ is the inverse matrix of
$\epsilon({\mathbf G-G'})$, $\perp$ and $\parallel$ refer to the
direction perpendicular and parallel to the $z$-axis. The Fourier
components $\epsilon({\mathbf G-G'})$ is calculated as \BE
\epsilon({\mathbf G}) = \frac{1}{A}\int_A \epsilon({\mathbf
r})e^{-i{\mathbf G}\cdot{\mathbf r}} d{\mathbf
r},\label{eq:ye7}\EE where the integration is performed over the
area of one lattice unit cell, and $A$ is the area.

For binary situations, Eq.~(\ref{eq:ye7}) can be further
simplified into \BE \epsilon({\mathbf G}) =
\left\{\begin{array}{ll} f\epsilon_a + (1-f)\epsilon_b &
\mbox{for} \ {\mathbf G} = 0, \\ (\epsilon_a-\epsilon_b)
S({\mathbf G}) & \mbox{for} \ {\mathbf G} \neq
0,\end{array}\right.\label{eq:ye8}\EE with $\epsilon_a$ and
$\epsilon_b$ referring to the dielectric constants for the
cylinders and background separately, and $f$ is the filling factor
defined as fraction of the area occupied by the cylinders in one
unit cell. The factor $S({\mathbf G})$ relies only one the
geometry of the cylinders and the lattice structures, and is given
by \beq S({\mathbf G})=\frac{1}{A}\int_{A_d} e^{-i{\mathbf
G}\cdot{\mathbf r}}d{\mathbf r}, \label{integral}\eeq where the
integration is carried over the area occupied by the cylinders in
the unit cell. We name $S({\mathbf G})$ as the structure factor.

\section{Evaluation of structure factors}

It is clear from the above derivation that once the structure
factor is known, the band structure for either the E-polarization
or H-polarization modes can be readily evaluated by taking the
structure factor into Eq.~(\ref{eq:ye8}) and subsequently into
Eqs.~(\ref{eq:E}) and (\ref{eq:H}) respectively. In the following
we evaluate the structure factor for common crystal structures.

\subsection{Transformation of $S({\mathbf G})$ under some
operations}

To proceed, first we discuss how the structure factor changes
under certain operations with regard to a single cylinder.
Different cylinders in a unit cell may undergo different
operations. The final structure factor of the unit cell will be
the sum of each individual structure factor. These operations
include {\it translation}, {\it rotation}, {\it reflection} and
{\it scaling} ({\it dilation} and {\it contraction}). Under these
operations the new $S({\mathbf G})$ is related to the original one
by applying simple transformations. The knowledge of these
transformations are useful in the calculation of the structure
factor for various lattices.

\subsubsection{Translation}

In the $x-y$ plane, the translational operation is the simplest
operation and it transforms the coordinates by a constant
displacement ${\mathbf r}_0$ as $$\mbox{T}_t[{\mathbf r}] =
{\mathbf r}'={\mathbf r}+{\mathbf r}_0.$$ Under this operation,
the structure factor changes as \beq \mbox{T}_t[S({\mathbf G})] =
S'= e^{-i{\mathbf G}\cdot {\mathbf r}_0}S({\mathbf G}). \eeq

\subsubsection{Rotation}

This operation rotates the cylinders by an angle $\theta$. Since
the calculation of structure factor must be coordinate
independent, we can rotate the coordinate system by the same angle
$\theta$ to simplify the calculation. In the new coordinates, the
structure factor will be the same form as the original one.
However, in the new system the vector ${\mathbf G}$ is transformed
as \beq \mbox{T}_r\left(\begin{array}{c} G_x\\
G_y\end{array}\right) = \left(\begin{array}{cc} \cos\theta &
\sin\theta\\ -\sin\theta &
\cos\theta\end{array}\right)\left(\begin{array}{c} G_x\\
G_y\end{array}\right), \EE where $G_x$ and $G_y$ are the ${\mathbf
G}$ components in the original coordinate system. Therefore under
this operation, we have \beq \mbox{T}_{r}[S({\mathbf G})] =
S(\mbox{T}_{r}[{\mathbf G}]). \eeq

\subsubsection{Reflection}

The reflection operation is to reflect the system about a line.
This operation will not change the form of the structure factor.
Rather it just change the reciprocal vector. By analogy with the
rotation operation, under the reflection operation we have \beq
\mbox{T}_{ref}[S({\mathbf G})] = S(\mbox{T}_{ref}[{\mathbf G}]).
\eeq

\subsubsection{Scaling}

The scaling operation includes dilatation and contraction of the
coordinates. Consider the operation \BE
\mbox{T}_s\left(\begin{array}{c} x\\y \end{array}\right)
=\left(\begin{array}{c} x' \\ y'  \end{array}\right) =
\left(\begin{array}{cc} \lambda_x & 0 \\ 0 &
\lambda_y\end{array}\right)\left(\begin{array}{c} x\\y
\end{array}\right)
\EE Under this operation \BE \mbox{T}_s[{\mathbf G}\cdot{\mathbf
r}] = {\mathbf G}\cdot\mbox{T}_s[{\mathbf r}]=\mbox{T}_s[{\mathbf
G}]\cdot{\mathbf r}, \label{scale1}\EE and \beq dA\rightarrow
\lambda_x\lambda_y dA.\label{scale2} \eeq From (\ref{scale1}) and
(\ref{scale2}) we find \beq \mbox{T}_s[S({\mathbf G})] =
 \lambda_x\lambda_y S(\mbox{T}_s[{\mathbf G}]).
\eeq

\subsubsection{The structure factors of circular and elliptic
cylinders}

As an example, we calculate the structure factors of a circular
and an elliptic cylinder. The latter is considered as a result of
the scaling operation on the former. For a circular cylinder of
radius $a$, we have  \beq S({\mathbf G}) = \frac{2\pi a
J_1(Ga)}{AG}. \eeq

For the elliptic rod, suppose the lengths of the principal axes
(along $x$ and $y$ axes respectively) are $a$ and $b$. According
to the idea before, we can obtain the result by a scaling
operation on the circular case. Here \beq
\lambda_x=1,\;\;\;\lambda_y=\frac{b}{a}, \eeq so \beq S({\mathbf
G}) =\frac{2\pi b J_1(\tilde{G}a)}{A\tilde{G}}, \eeq where \beq
\tilde{G}=\sqrt{G^2_x+(\frac{b}{a})^2G^2_y}. \eeq

\subsection{Structure factor of a polygonal cylinder}

Now we consider the general case. Consider a polygonal cylinder
with $N$ sides. The $N$ corners of the polygon are labeled as
${\mathbf r}_{1},{\mathbf r}_{2}, \ldots,{\mathbf r}_{N}$, where
the $j$th corner ${\mathbf r}_j=(x_j,y_j)$, and we define
${\mathbf r}_{N+1}={\mathbf r}_1$. Suppose we can find a vector
field ${\mathbf F}$ such that \beq (\nabla\times {\mathbf
F})_z=e^{-i{\mathbf G}\cdot{\mathbf r}}, \eeq then according to
Stokes' Theorem, we have \beq \int_{A_d} \nabla\times {\mathbf
F}\cdot d{\mathbf A} =\oint_{C_d} {\mathbf F}\cdot d{\mathbf
r},\;\;d{\mathbf A}=dA \hat{z}, \eeq and thus the original
integral (\ref{integral}) is reduced to computing the line
integral \beq I=\oint_{C_d} {\mathbf F}\cdot d{\mathbf
r},\label{ctr} \eeq where $C_d$ is the boundary of the polygon. In
the rest we present a detailed calculation of this line
integration.

To begin with, assume $G_x\neq 0$. Choose \beq {\mathbf F}=F_y
\hat{y},\label{f1} \eeq then solving the equation
\[
(\nabla\times {\mathbf F})_z=\partial_x F_y
=e^{-i{\mathbf G}\cdot {\mathbf r}}=e^{-iG_x x}e^{-iG_y y}
\]
will gives us the solution \beq F_y=\frac{i}{G_x}e^{-i{\mathbf
G}\cdot{\mathbf r}}.\label{f1a} \eeq The integral $I$ now becomes
\beq
I=\frac{i}{G_x}\sum^N_{j=1}\int^{(x_{j+1},y_{j+1})}_{(x_j,y_j)}
e^{-i(G_xx+G_yy)}dy,\label{ieg} \eeq where
$\int^{(x_{j+1},y_{j+1})}_{(x_j,y_j)}$ denotes the line integral
between $j$-th and $(j+1)$-th corner, i.e., the $j$-th side. For
the $j$-th side we have\bqn
&&\int^{(x_{j+1},y_{j+1})}_{(x_j,y_j)}e^{-i(G_xx+G_yy)}dy\non\\
&=&\frac{i(y_{j+1}-y_j) (e^{-i{\mathbf G}\cdot {\mathbf r}_{j+1}}
-e^{-i{\mathbf G}\cdot {\mathbf r}_j })} {{\mathbf
G}\cdot({\mathbf r}_{j+1}-{\mathbf r}_j)}\non\\ &=&(y_{j+1}-y_j)
e^{-i{\mathbf G}\cdot (\frac{{\mathbf r}_{j+1}+{\mathbf r}_j}{2})}
\frac{\sin[{\mathbf G}\cdot (\frac{{\mathbf r}_{j+1}-{\mathbf
r}_j}{2})]} {{\mathbf G}\cdot(\frac{{\mathbf r}_{j+1}-{\mathbf
r}_j}{2})}. \label{der} \eqn Substituting (\ref{der}) into
(\ref{ieg}) and defining \beq {\mathbf
C}_j\equiv\left(\frac{{\mathbf r}_{j+1}+{\mathbf
r}_j}{2}\right),\;\; {\mathbf S}_j\equiv\left(\frac{{\mathbf
r}_{j+1}-{\mathbf r}_j}{2}\right), \eeq \beq \Delta y_j\equiv
y_{j+1}-y_j, \eeq we find \beq I=\sum^N_{j=1}\frac{i\Delta y_j
e^{-i{\mathbf G}\cdot {\mathbf C}_j}}{G_x}
\frac{\sin\left({\mathbf G}\cdot {\mathbf S}_j\right)} {{\mathbf
G}\cdot{\mathbf S}_j}.\label{result} \eeq

When $G_x=0$, Eq.(\ref{result}) no longer holds. We have to find
another expression. In this case, since $G_y\neq 0$, we can choose
\beq {\mathbf F}=F_x\hat{x}\label{f2} \eeq and solve
\[
(\nabla\times {\mathbf F})_z=-\partial_yF_x=e^{-iG_xx}e^{-iG_yy}
\]
to find the solution \beq F_x=-\frac{i}{G_y}e^{-i{G}\cdot{\mathbf
r}}.\label{f2a} \eeq Taking this ${\mathbf F}$ into Eq.(\ref{ctr})
and defining \beq \Delta x_j\equiv x_{j+1}-x_j, \eeq we find \beq
I=\sum^N_{j=1}\frac{-i\Delta x_j e^{-i{\mathbf G}\cdot {\mathbf
C}_j}}{G_y} \frac{\sin\left({\mathbf G}\cdot {\mathbf S}_j\right)}
{{\mathbf G}\cdot{\mathbf S}_j}. \label{result1} \eeq

In fact, both (\ref{result}) and (\ref{result1}) can be written in
the unified expression \bqn I&=&\sum^N_{j=1}\frac{2i{\mathbf
S}_j\cdot\hat{n}_2 e^{-i{\mathbf G}\cdot {\mathbf C}_j}} {{\mathbf
G}\cdot \hat{n}_1} \frac{\sin\left({\mathbf G}\cdot {\mathbf
S}_j\right)} {{\mathbf G}\cdot{\mathbf S}_j}\non\\
&=&\sum^N_{j=1}\frac{2i\hat{z}\cdot(\hat{n}_1\times{\mathbf S}_j)
e^{-i{\mathbf G}\cdot {\mathbf C}_j}} {{\mathbf G}\cdot \hat{n}_1}
\frac{\sin\left({\mathbf G}\cdot {\mathbf S}_j\right)} {{\mathbf
G}\cdot{\mathbf S}_j},\label{unify} \eqn where $\hat{n}_1$ is an
arbitrary unit vector, and $\hat{n}_2$ is another unit vector
defined by $\hat{n}_2=\hat{z}\times \hat{n}_1$.

If we choose $\hat{n}_1$ as \beq \hat{n}_1=\frac{\mathbf G}{G},
\eeq where $G=|\mathbf G|$, we then obtain a general expression
\beq I=\sum^N_{j=1}\frac{2i\hat{z}\cdot({\mathbf G}\times{\mathbf
S}_j) e^{-i{\mathbf G}\cdot{\mathbf C}_j}}{G^2}
\frac{\sin\left({\mathbf G}\cdot {\mathbf S}_j\right)} {{\mathbf
G}\cdot{\mathbf S}_j},\label{cordindp} \eeq for either $G_x \neq
0$ or $G_y\neq 0$.

In the above we have obtained the general expression
Eq.~(\ref{cordindp}) of the integral $I$ when ${\mathbf G}\neq 0$.
Several remarks are worthwhile.

\begin{enumerate}

\item If ${\mathbf G}=0$, the integral
in (\ref{integral}) is simply $A$, i.e., the cross section area.
However, we do not need to consider this situation because the
structure factor at ${\mathbf G}=0$ does not involve in the band
structure calculation, referring to Eq.~(\ref{eq:ye8}).

\item Formula (\ref{result}),
(\ref{result1}), (\ref{unify}) and (\ref{cordindp}) are still
correct when the polygon region has several polygons. In this case
the sum in (\ref{cordindp}) can be rewritten as \beq
I=\sum^P_{p=1}\sum^{N_p}_{j_p=1}\frac{2i\hat{z}\cdot({\mathbf
G}\times {\mathbf S}_{j_p}) e^{-i{\mathbf G}\cdot{\mathbf
C}_{j_p}}}{G^2} \frac{\sin\left({\mathbf G}\cdot {\mathbf
S}_{j_p}\right)} {{\mathbf G}\cdot{\mathbf S}_{j_p}},\label{large}
\eeq where $p=1,\ldots,P$ and $P$ is the total number of polygons;
$j_p=1,\ldots,N_p$, with $N_p$ being the total number of corners
of the $p$-th polygon.

\item Since the integral (\ref{ctr}) is calculated along
a closed loop, we can always gauge transform the field ${\mathbf
F}$ by adding an arbitrary gradient: \beq {\mathbf F}'={\mathbf
F}+\nabla \Lambda, \eeq where $\Lambda$ is a scalar field. Under
this gauge transformation, the structure factor is not changed.
For example, denote ${\mathbf F}$ in (\ref{f1a}) as ${\mathbf
F}_1$ and the ${\mathbf F}$ in (\ref{f2a}) as ${\mathbf F}_2$,
then \beq {\mathbf F}_1-{\mathbf F}_2 =\nabla\left(
\frac{-e^{-i{\mathbf G}\cdot{\mathbf r}}}{G_xG_y}\right). \eeq
Therefore ${\mathbf F}_1$ and ${\mathbf F}_2$ differs by a
gradient, and both gave the same structure factor.

\item If $G_x=0$, one can use Eq.(\ref{result1}); if $G_y=0$,
one can use Eq. (\ref{result}). It can be proved that the result
with $G_x=0$ can also be obtained from Eq. (\ref{result}) by
taking the $G_x\rightarrow 0$ limit. Similarly, when $G_y=0$, one
can obtain the correct result from Eq. (\ref{result}) by taking
the $G_y\rightarrow 0$ limit.

\item For later convenience, we define a function $Q(x)$ as
\beq Q(x)\equiv \frac{\sin x}{x}. \eeq

\end{enumerate}

\section{Some analytic examples}

Now we apply the above formulas to calculate explicitly the
structure factors of various configurations.

\subsection{Square lattice}

First we consider the square lattice. The lattice constant is
supposed to be $d$. The base vectors are \beq {\mathbf
a}_1=d\hat{x},\;\;{\mathbf a}_2=d\hat{y}, \eeq so the area of a
unit cell is \beq A=d^2. \eeq The bases of the reciprocal lattice
are \beq {\mathbf b}_1=\frac{2\pi}{d}\hat{x},\;\;{\mathbf
b}_2=\frac{2\pi}{d}\hat{y}, \eeq which give \beq {\mathbf
G}=n_1{\mathbf b}_1+n_2{\mathbf b}_2
=\frac{2\pi}{d}\left(n_1\hat{x}+n_2\hat{y}\right), \eeq or \beq
G_x=\frac{2\pi n_1}{d},\;\;G_y=\frac{2\pi n_2}{d}. \eeq

\subsubsection{Square lattice with circular cylinders (SC)}

First we arrange the circular cylinders in the square lattice.
This is simplest type of photonic crystals\cite{Plihal}. The
radius of the cylinder is $a$. The structure factor is given by
\beq S({\mathbf G}) = 2f\frac{J_1(Ga)}{Ga}. \eeq The filling
fraction $f$ and its maximum value $f_{max}$ are \beq f=\frac{\pi
a^2}{A}=\pi\left(\frac{a}{d}\right)^2,\;\; f_{max}=\frac{\pi}{4}.
\eeq For a general $(n_1,n_2)$ pair we have \beq
Ga=\sqrt{\frac{4\pi^2 a^2}{d^2}(n^2_1+n^2_2)}=\sqrt{4\pi
f(n^2_1+n^2_2)}. \eeq

\subsubsection{Square lattice with square cylinders (SS)}

Now we place the square cylinders on the square
lattice\cite{noncircular}. The side length of the square is $a$.
The structure factor is given by \BE S({\mathbf G}) =
fQ\left(\frac{G_xa}{2}\right)Q\left(\frac{G_ya}{2}\right) =
fQ\left(\frac{n_1\pi a}{d}\right)Q\left(\frac{n_2\pi a}{d}\right).
\EE Here the filling fraction and its maximum value are \beq
f=\left(\frac{a}{d}\right)^2,\;\; f_{max}=1. \eeq

\subsection{Triangular lattice}

In this case we still assume the lattice constant $l$ and circle
radius $a$. The base vectors are given by \beq {\mathbf
a}_1=d\hat{x},\;\;{\mathbf
a}_2=d(\frac{1}{2}\hat{x}+\frac{\sqrt{3}}{2}\hat{y}), \eeq so the
area of a unit cell is \beq A=\frac{\sqrt{3}}{2}\,d^2. \eeq The
bases of the reciprocal lattice are \beq {\mathbf
b}_1=\frac{2\pi}{d}(\hat{x}-\frac{1}{\sqrt{3}}\hat{y}),
\;\;{\mathbf b}_2=\frac{2\pi}{d}\frac{2}{\sqrt{3}}\hat{y}, \eeq
which give \beq {\mathbf G}=n_1{\mathbf b}_1+n_2{\mathbf b}_2
=\frac{2\pi}{d}\left(n_1\hat{x}+
\frac{2n_2-n_1}{\sqrt{3}}\hat{y}\right), \eeq or \beq
G_x=\frac{2\pi n_1}{d},\;\;
G_y=\frac{2\pi}{d}\left(\frac{2n_2-n_1}{\sqrt{3}}\right). \eeq

\subsubsection{Triangular lattice with circular cylinder (TC)}

This is the case considered by \cite{Plihal2}. The radius of the
circle is $a$. The structure factor is \beq
S=\frac{I}{A}=2f\frac{J_1(Ga)}{Ga}, \eeq with \beq f=\frac{\pi
a^2}{A} =\frac{2\pi}{\sqrt{3}}\left(\frac{a}{d}\right)^2,\;\;
f_{max}=\frac{\pi}{2\sqrt{3}}. \eeq For a general $(n_1,n_2)$ pair
we have \bqn Ga&=&\sqrt{\frac{16\pi^2
a^2}{3d^2}(n^2_1+n^2_2-n_1n_2)}\non\\ &=&\sqrt{\frac{8\pi
f}{\sqrt{3}}(n^2_1+n^2_2-n_1n_2)}. \eqn

\subsubsection{Triangular lattice with hexagonal cylinder (TH)}

In this case we assume the hexagon side length is $a$, thus the
cross section area is \beq A=\frac{3\sqrt{3}\,a^2}{2}. \eeq The
filling fraction is given by \beq
f=\frac{A}{A}=3\left(\frac{a}{d}\right)^2, \eeq with maximum value
\beq f_{max}=\frac{3}{4}. \eeq The structure factor is \BE
S({\mathbf G}) = \frac{(2f/3)}{G_xa}\left\{
\sin\!\left[\frac{(3G_x+\sqrt{3}G_y)a}{4}\right]\!
Q\!\left[\frac{(G_x-\sqrt{3}G_y)a}{4}\!\right] \right.
\left.+\sin\!\left[\frac{(3G_x-\sqrt{3}G_y)a}{4}\right]\!
Q\!\left[\frac{(G_x+\sqrt{3}G_y)a}{4}\right]\! \right\}\label{str}
\EE or \BE S({\mathbf G}) = \frac{(fd/3)}{n_1\pi a} \left\{
\sin\left[\frac{(n_1+n_2)\pi a}{d}\right]
Q\left[\frac{(n_1-n_2)\pi a}{d}\right]\right.\left.
+\sin\left[\frac{(2n_1-n_2)\pi a}{d}\right] Q\left[\frac{ n_2\pi
a}{d}\right] \right\}.\label{str1} \EE

\subsection{Tables of structure factors}

\begin{figure}[hbt]
\begin{center}
\epsfxsize=4.5in \epsffile{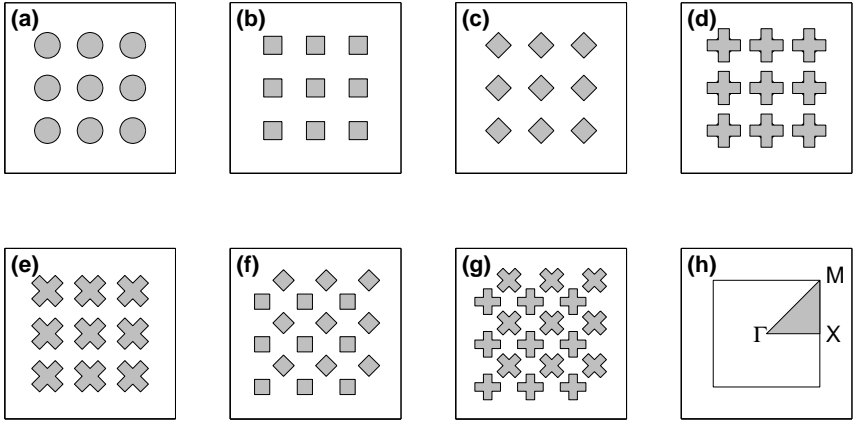}\vspace{5mm}
\caption{\label{figure1}\small Patterns of the square lattice. (a)
square-circle (SC). (b) square-square (SS). (c) square-rotated
square (SRS). (d) square-cross (SCR). (e) square-rotated cross
(SRCR). (f) square-mixed square (SMS). (g) square-mixed cross
(SMCR). (h) The first Brillouin zone of square lattice. The band
structure is calculated along the boundary $\mbox{M}-\Gamma-
\mbox{X}-\mbox{M}$ of the gray region.}
\end{center}
\end{figure}

\begin{figure}[hbt]
\begin{center}
\epsfxsize=4.5in \epsffile{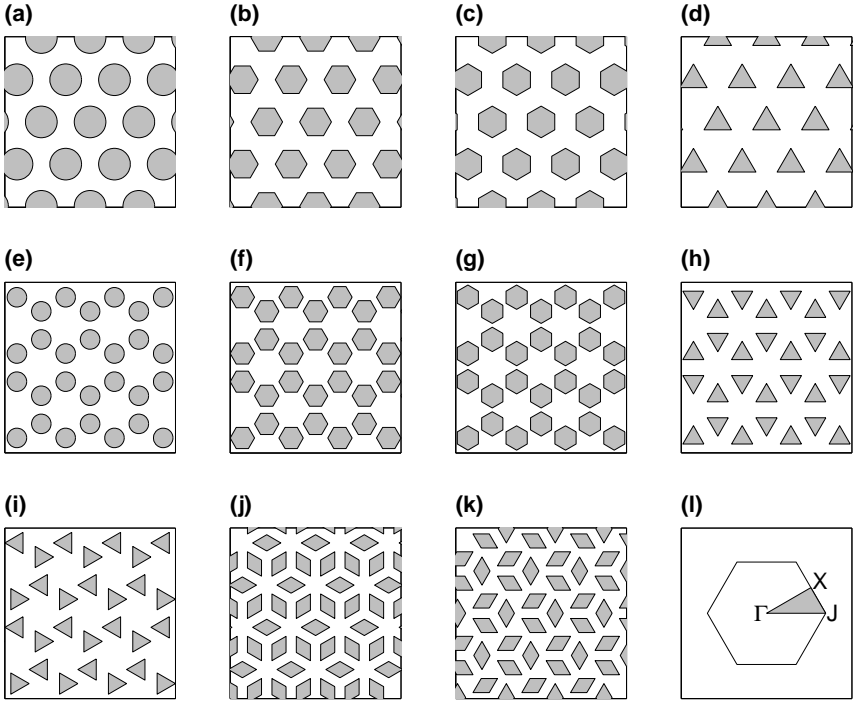}\vspace{5mm}
\caption{\label{figure2}\small Patterns of the triangular lattice.
(a) triangular-circle (TC). (b) triangular-hexagon (TH). (c)
triangular-rotated hexagon (TRH). (d) triangular-triangle (TT).
(e) honeycomb-circle (HC). (f) honeycomb-hexagon (HH). (g)
honeycomb-rotated hexagon (HRH). (h) honeycomb-triangle (HT). (i)
honeycomb-rotated triangle (HRT). (j) triangular-diamonds (TD).
(k) triangular-rotated diamonds (TRD). (l) The first Brillouin
zone of triangular lattice. The band structure is calculated along
the boundary $\mbox{X}-\Gamma- \mbox{J}-\mbox{X}$ of the gray
region.}
\end{center}
\end{figure}

In this section we consider various arrangements of rod cross
section and lattices. Eighteen patterns (configurations) are
illustrated in Fig.~(\ref{figure1}) and (\ref{figure2}). Here we
consider square, triangular, and honeycomb lattices. The
abbreviation of these patterns are defined in the figure captions.
For example, TC refers to the situation that circular cylinders
are arranged in a triangular lattice. For a circular rod the
parameter $a$ denotes the radius of cylinder. For a polygonal
cylinder, $a$ represents the side length. For a ``cross"-shaped or
a ``rotated cross"-shaped rod, $a$ and $\alpha a$ represent the
``long" and ``short" side lengths of the corresponding rectangle.
In a unit cell of the ``triangular-diamonds" or
``triangular-rotated diamonds" pattern (See Figs.~\ref{figure2}(j)
and (k)) there are three ``diamonds". In the ``rotated diamonds"
pattern the distance between the center of each diamond and the
center of unit cell is chosen to be $d/4$; $d$ is the lattice
constant.

The crystal structures shown by these two figures are most common
in photonic design. While some have already been published, many
are reported for the first time. In the next section, a new class
of photonic crystals will be investigated. The corresponding band
structure will be computed. The corresponding properties and the
forms of unit cells are given in Table I, and the structure
factors are listed in Table II, respectively. In these tables, the
first column refers to the combination of the lattice structure
and the shape of rods, the second to the unit cells, the third to
the filling factor formula, with the maximum value being listed in
the fourth column. The next two columns refer to the two base
reciprocal vectors.

\section{The numerical results}

In this section we consider the band structure of the 2D photonic
crystals made by drilling hollow cross-shaped cylinders in an
alumina ceramic background and by placing cross-shaped alumina
ceramic cylinders in the air. The cross-shaped cylinders are
arranged in a square lattice configuration, corresponding to
Fig.~\ref{figure1}(d). The following parameters are used: filling
factor $f=0.5$, the ratio between $b$ and $a$ is 0.3. The
dielectric constant of the alumina ceramic is 8.9. We used more
than 600 plane waves. The inaccuracy is less than 1\%.

\begin{figure}[hbt]
\begin{center}
\epsfxsize=5in \epsffile{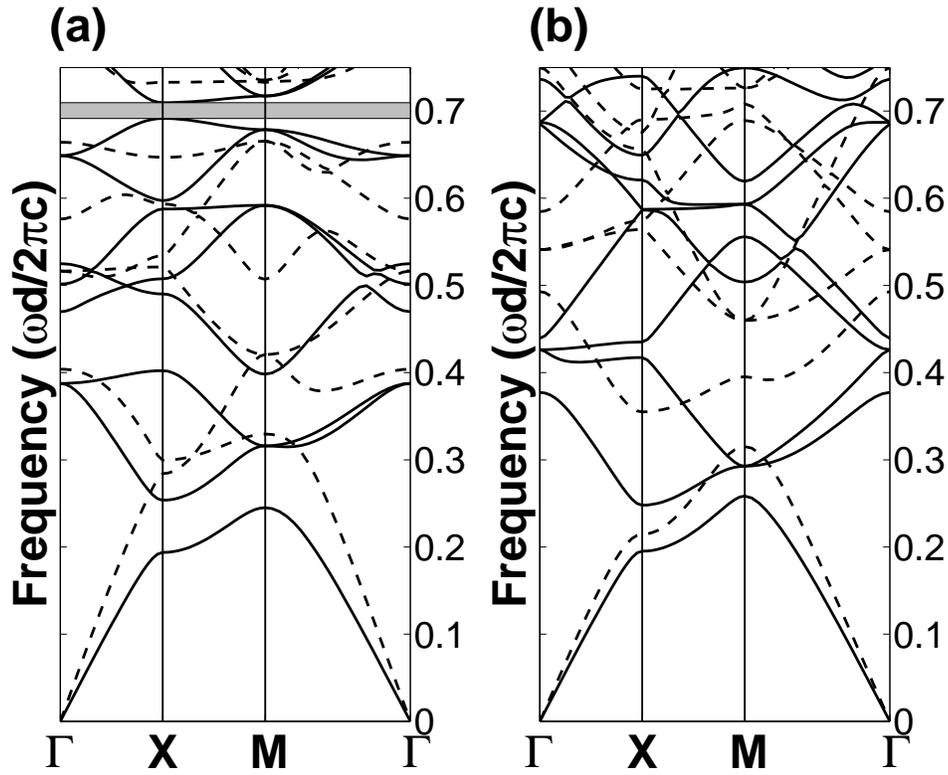}\vspace{5mm}
\caption{\label{figuregap}\small The photonic band structures. The
solid and dashed lines refer to the E-polarization and
H-polarization modes respectively. (a) Air cylinders in alumina
ceramics. The shaded area denotes the complete band gap. (b)
Alumina ceramics cylinders in air.}
\end{center}
\end{figure}

The photonic band structures are plotted in Fig.~\ref{figuregap}.
Here we see from (a) that an absolute band gap appears in the high
frequency range for the E- and H- polarizations, in the case of
air cylinders in the dielectric medium. Incidently, the absolute
band gap position is at around the same regime as in \cite{korea}
and \cite{brief}; we note that in \cite{korea}, it was the very
high dielectric cylinders that are placed in air. This indicates
that it is relatively easy to obtain the absolute band gap in the
infrared regime with this type of photonic crystals. We also see
that there are more complete gaps in the E-polarization mode than
in the H-mode. These occur at around $\omega d/(2\pi c) = 0.25,
0.6$, but the largest gap happens at the position of the absolute
gap around $\omega d/(2\pi c) = 0.7$, where the gap of the
H-polarization is wider than that of the E-polarization. In the
opposite case where the cylinders are made of high dielectric
constant materials, there is no absolute band gap at all. In fact,
there is no gap for the E-polarization. However, a complete gap
appears around $\omega d/(2\pi c) = 0.35$ for the H-polarization.

\section{Summary}

In this article, we derived necessary formulas needed for
computing the band structures of various types of photonic
crystals. The results are systematically presented in tables, so
that the reader could easily make a use of them. As an example, we
applied the results to two dimensional arrays of hollow
cross-shaped cylinders embedded in an alumina ceramic medium. For
this new class of photonic crystals, a large absolute phontonic
band gap is discovered in the high frequency region.

Finally, we want to make a connection between the photonic and
sonic crystals. In the acoustic case, the wave equation is
governed by \BE \nabla\cdot\left(\frac{1}{\rho({\mathbf r})}
\nabla p({\mathbf r})\right) + \frac{\omega^2}{\rho({\mathbf
r})c^2({\mathbf r})} p({\mathbf r})=0. \label{eq:sound}\EE In the
H-polarization mode, Maxwell equation is from Eq.~(\ref{eq:ye1})
\BE \nabla\cdot\left(\frac{1}{\epsilon({\mathbf r})} \nabla
H_{\parallel}({\mathbf r})\right) +
\frac{\omega^2}{\epsilon({\mathbf r})c^2({\mathbf r})}
H_{\parallel}({\mathbf r})=0,\EE where $c^2 =
c_0^2/\epsilon({\mathbf r})$. Therefore the acoustic pressure
field and the magnetic field have an one-to-one correspondence
given by the following mapping, \BE \rho({\mathbf r})
\rightleftharpoons \epsilon({\mathbf r}), \ \ \ \rho({\mathbf
r})c^2({\mathbf r}) \rightleftharpoons \epsilon({\mathbf
r})c^2({\mathbf r}). \EE

\section*{Acknowledgments}

This work received support from the National Science Council.

\input{epsf}
\begin{center}
\end{center}

\input{epsf}
\begin{center}
\end{center}

\input{epsf}
\begin{center}
\end{center}




\onecolumn
\begin{table}
\label{tab1}
\caption{Properties of various patterns}
\begin{tabular}{llllll}
{\Large $\Box$} latt. &Unit cell&$f$&$f_{max}$&$G_x$&$G_y$\\
\hline
SC&\epsfxsize=0.9cm \epsffile{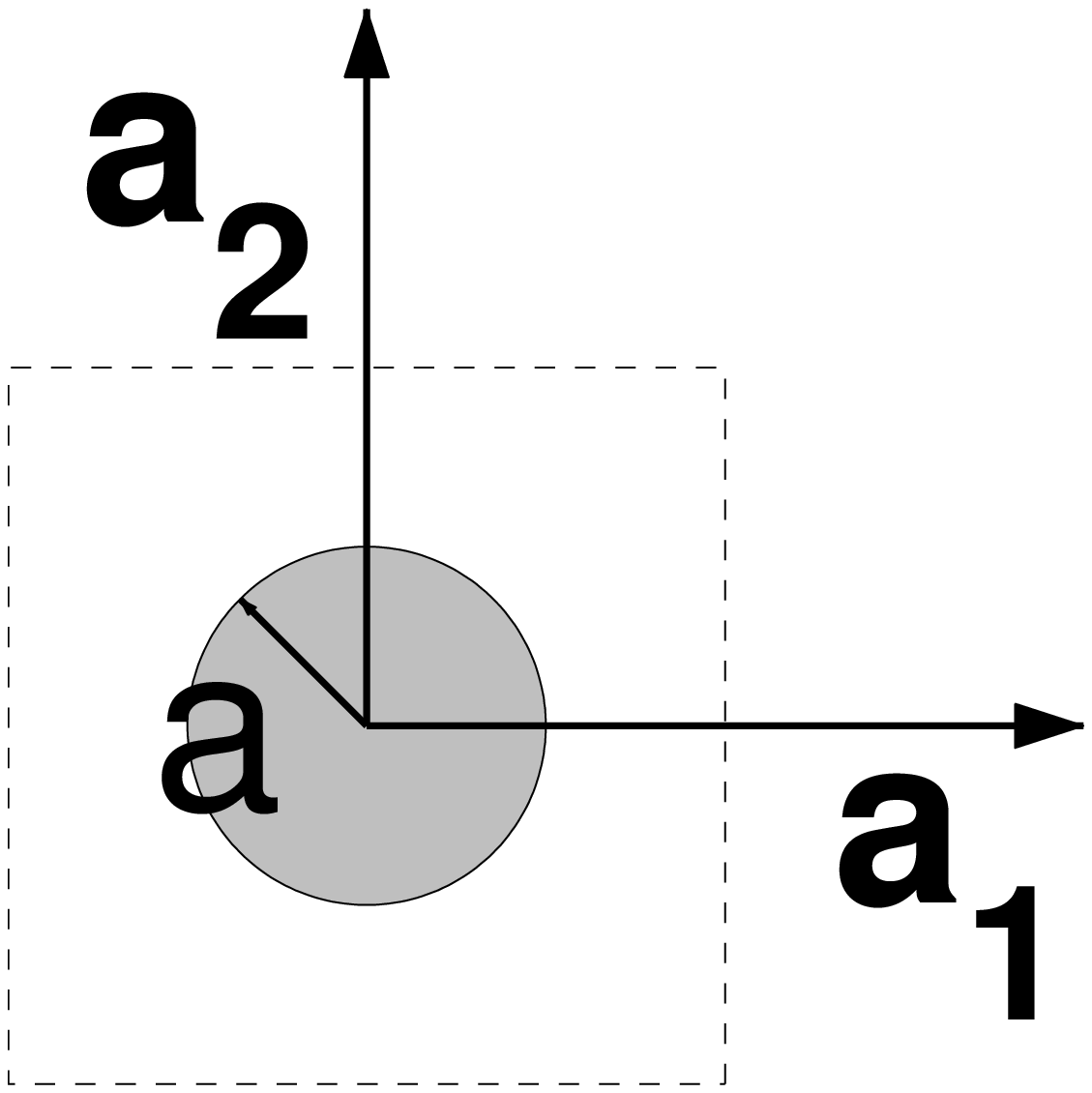}&$\frac{\pi a^2}{d^2}$ &$\frac{\pi}{4}$&$\frac{2\pi n_1}{d}$&$\frac{2\pi n_2}{d}$\\
SS&\epsfxsize=0.9cm\epsffile{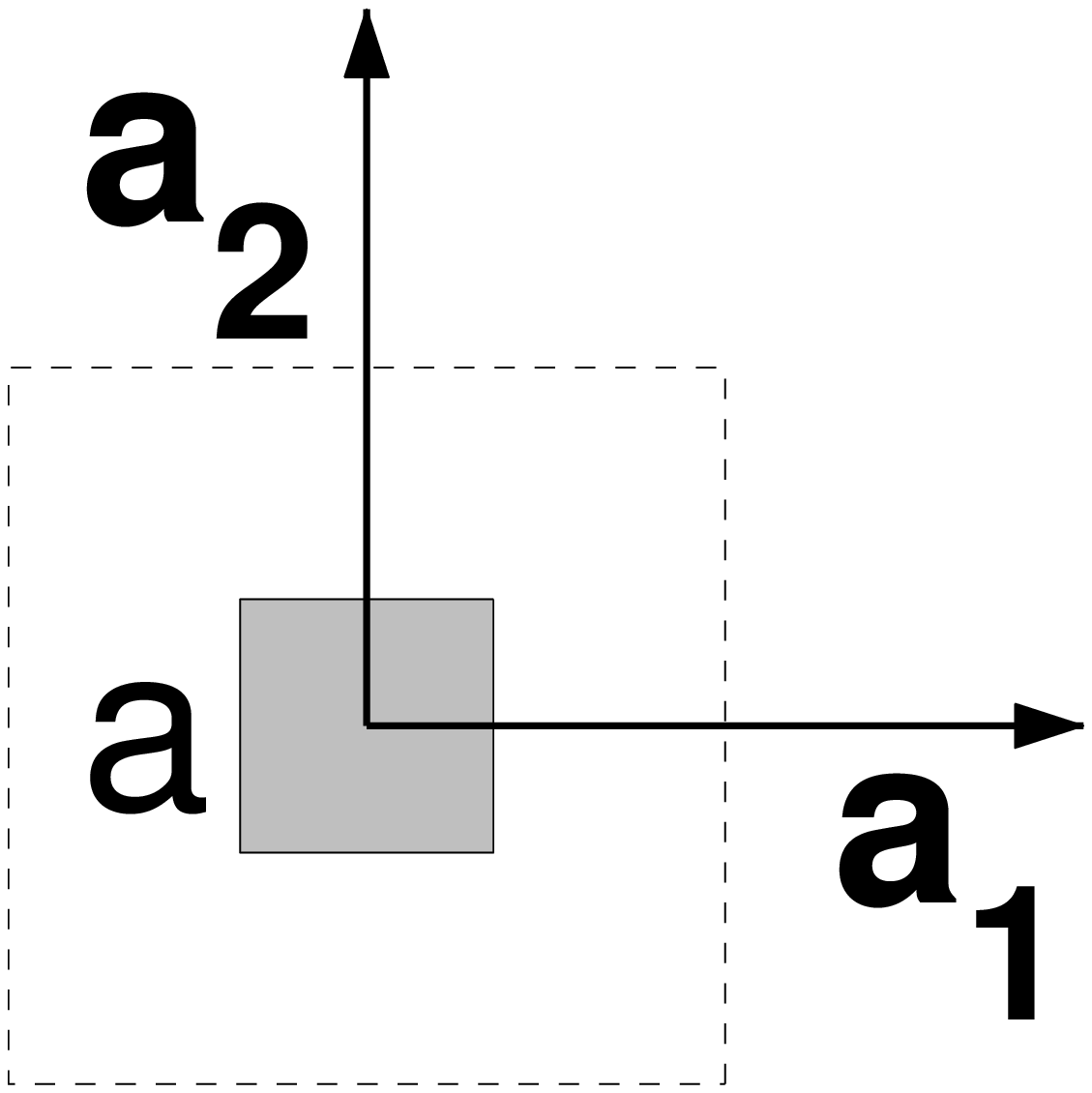}&$\frac{a^2}{d^2}$&$1$&$\frac{2\pi n_1}{d}$&$\frac{2\pi n_2}{d}$\\
SRS&\epsfxsize=0.9cm\epsffile{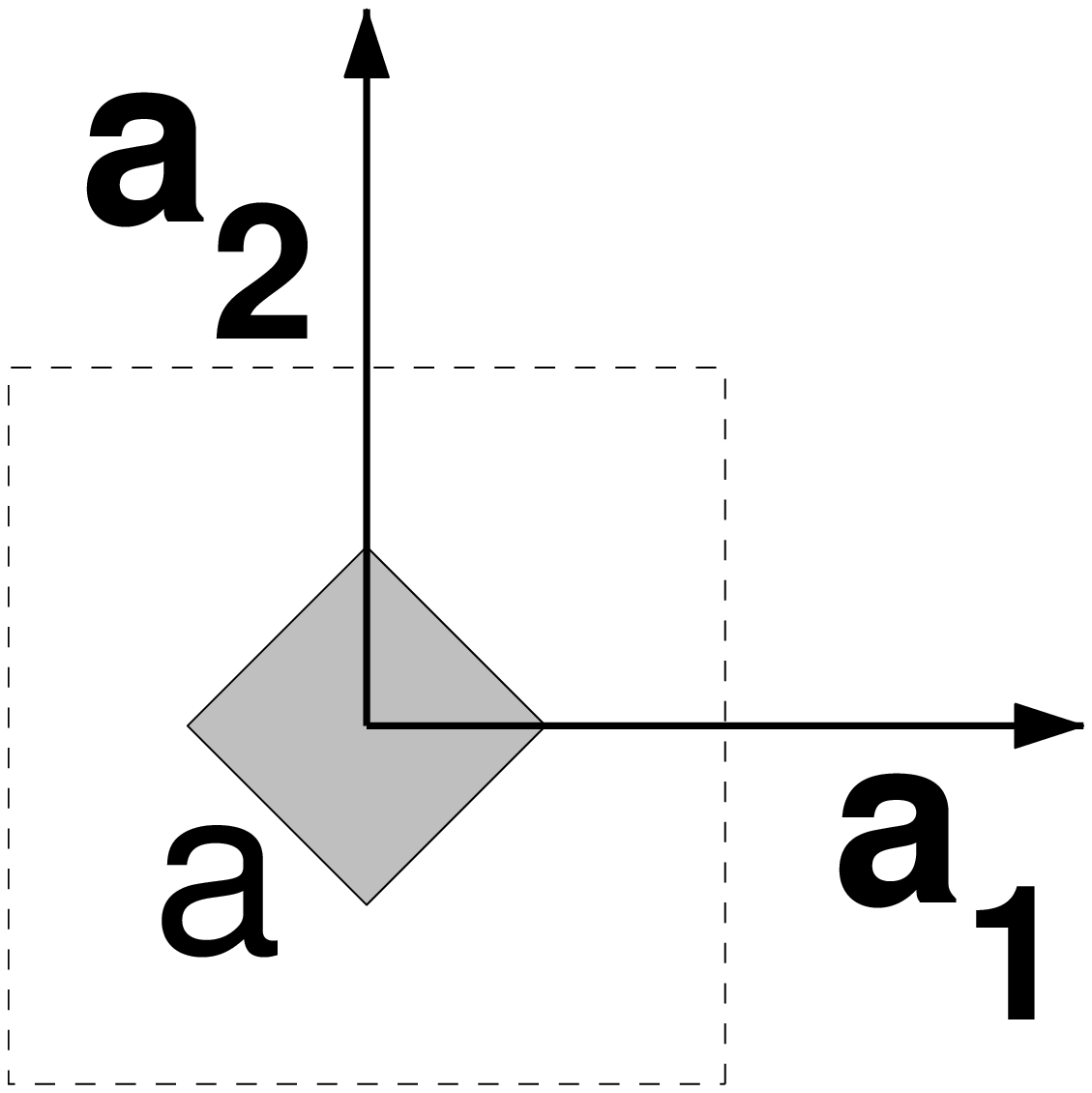}&$\frac{a^2}{d^2}$&$\frac{1}{2}$&$\frac{2\pi n_1}{d}$&$\frac{2\pi n_2}{d}$\\
SCR&\epsfxsize=1cm\epsffile{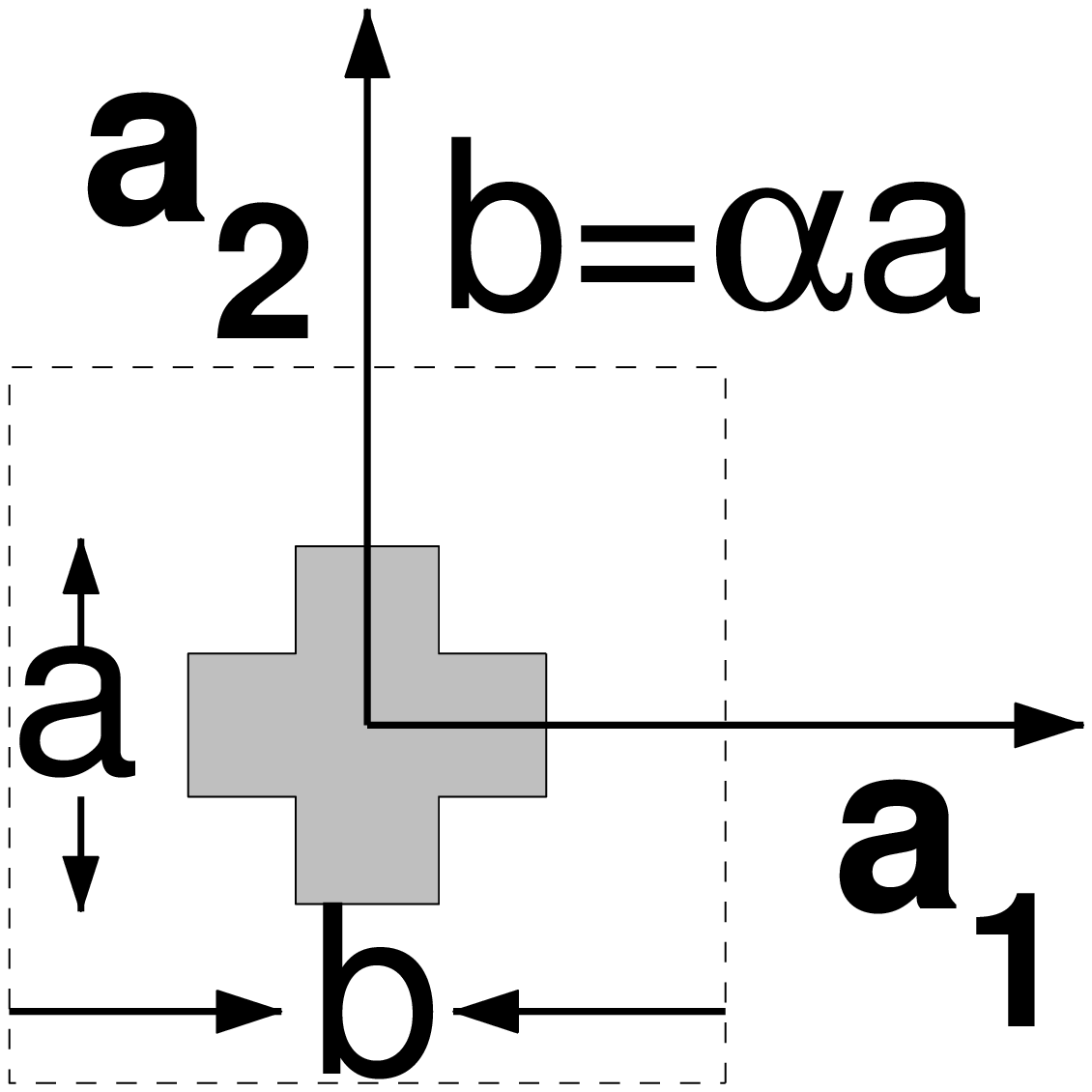}&$\frac{\alpha(2-\alpha)a^2}{d^2}$&
$\alpha(2-\alpha)$&$\frac{2\pi n_1}{d}$&$\frac{2\pi n_2}{d}$\\
SRCR&\epsfxsize=0.9cm \epsffile{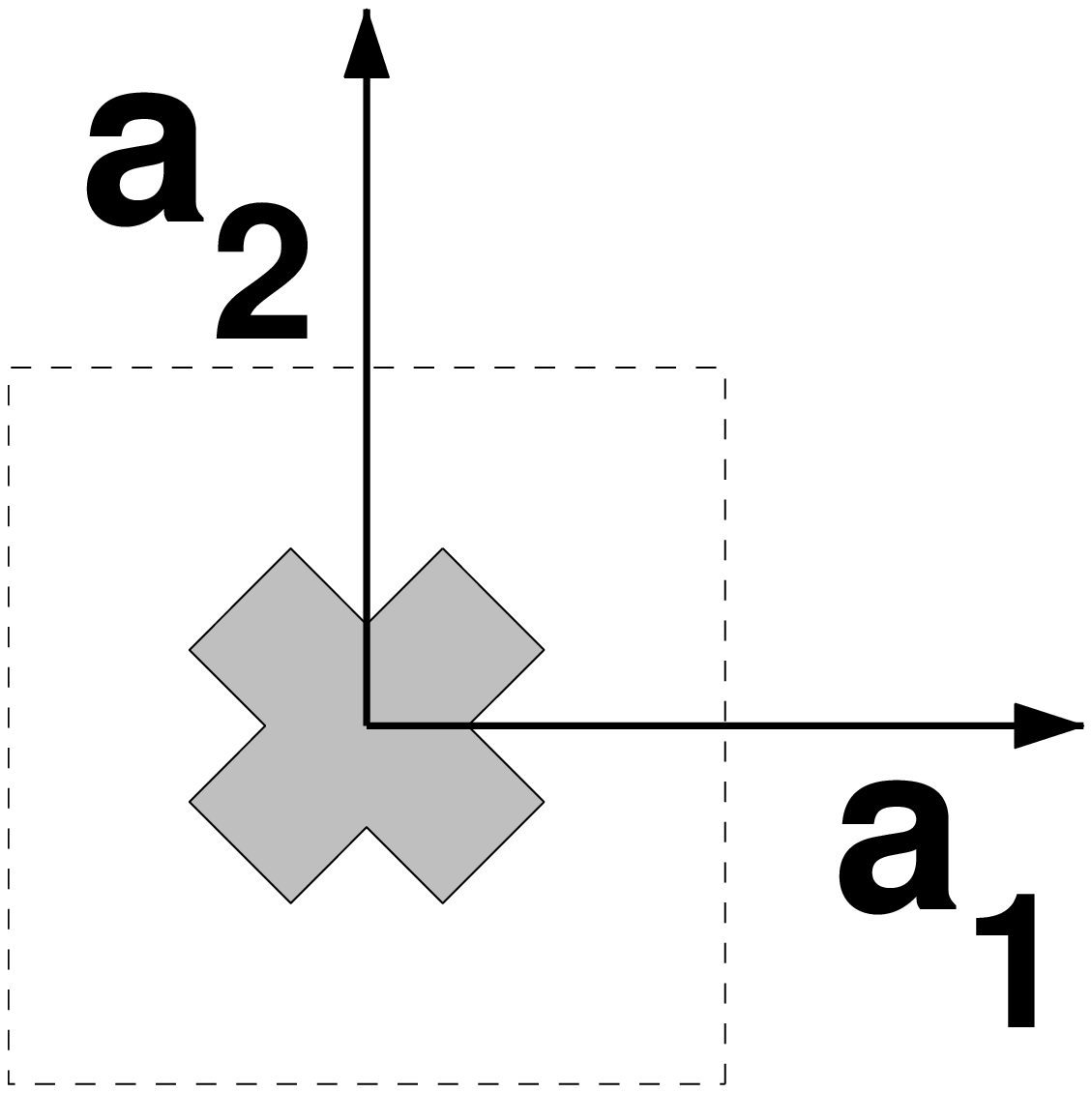}&$\frac{\alpha(2-\alpha)a^2}{d^2}$
&$\frac{2\alpha(2-\alpha)}{(1+\alpha)^2}$&$\frac{2\pi n_1}{d}$&$\frac{2\pi n_2}{d}$\\
SMS&\epsfxsize=0.9cm\epsffile{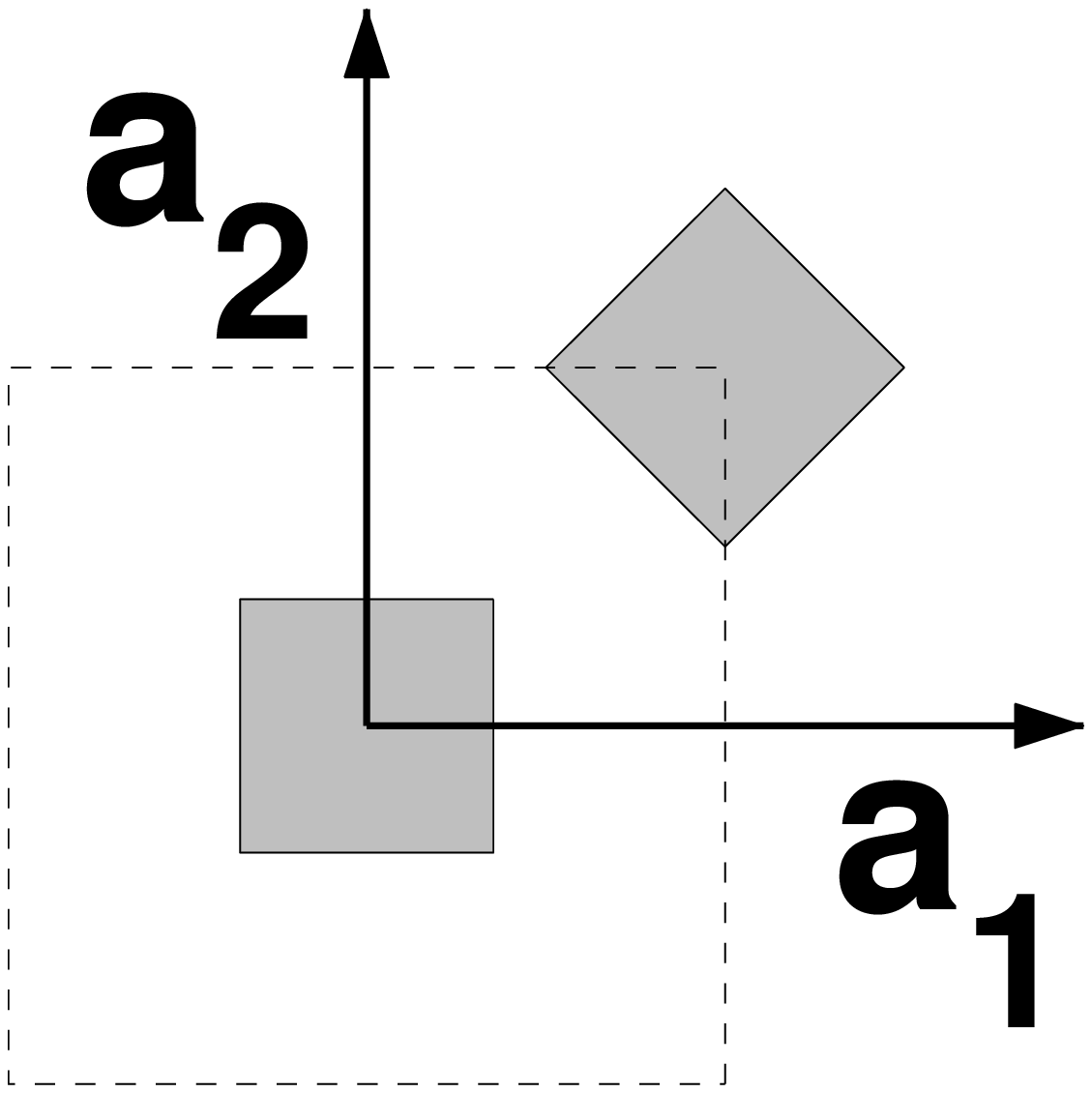}&$\frac{2a^2}{d^2}$&$4(3-2\sqrt{2})$
&$\frac{2\pi n_1}{d}$&$\frac{2\pi n_2}{d}$\\
SMCR&\epsfxsize=0.9cm\epsffile{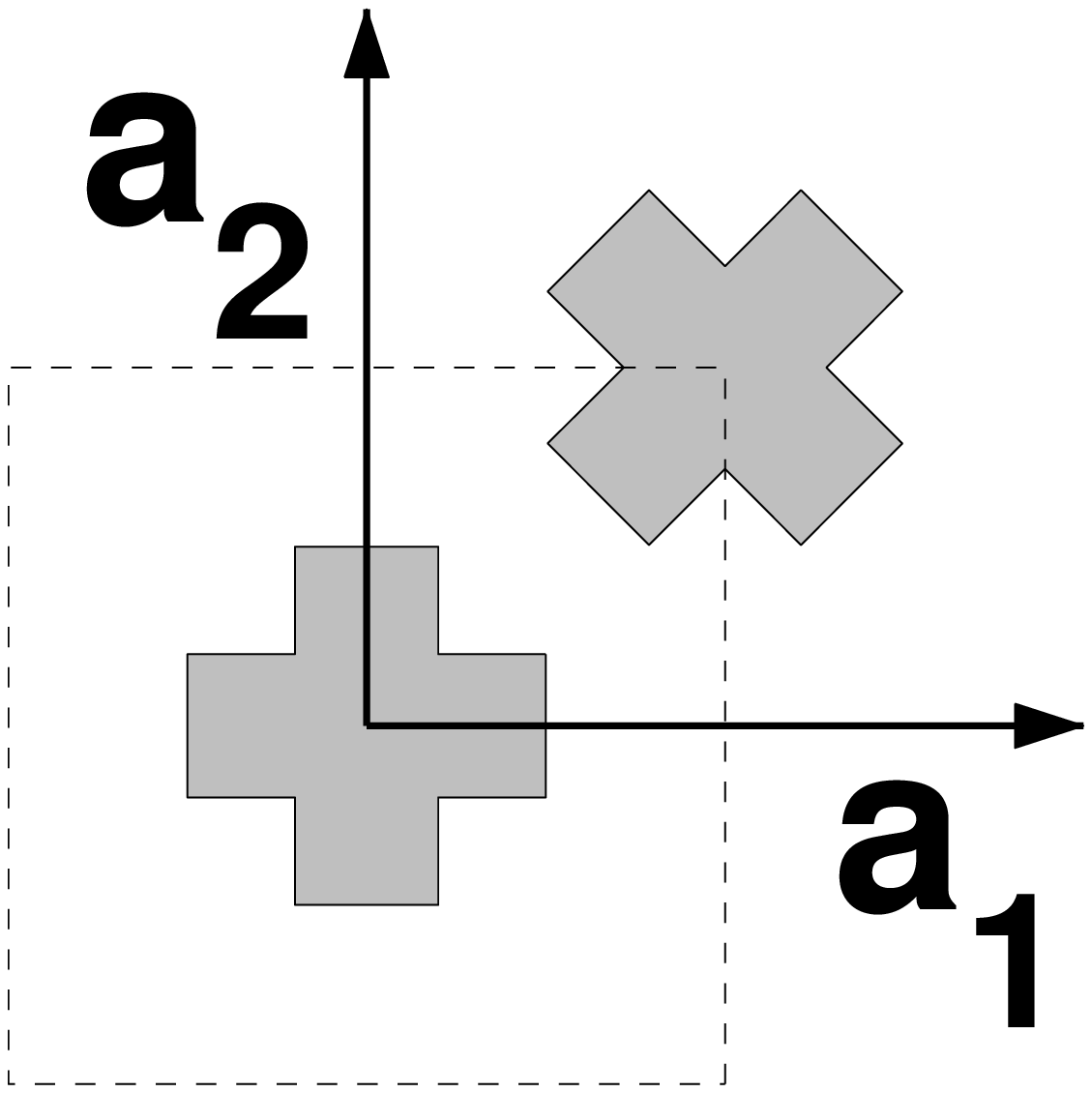}&$\frac{2\alpha(2-\alpha)a^2}{d^2}$&
$\frac{4\alpha(2-\alpha)}{[(\sqrt{2}+1)\alpha+1]^2}\;
\;\mbox{if}\;\alpha\leq\sqrt{2}-1$&$\frac{2\pi n_1}{d}$&
$\frac{2\pi n_2}{d}$\\
&&&$\frac{8\alpha(2-\alpha)}{[\alpha+(\sqrt{2}+1)]^2}\;\;\;\mbox{if}\;\alpha>\sqrt{2}-1$&&\\
&&&&&\\
\hline\\
{\large$\triangle$} latt. &Unit cell&$f$&$f_{max}$&$G_x$&$G_y$\\
\hline\\
TC&\epsfxsize=0.95cm\epsffile{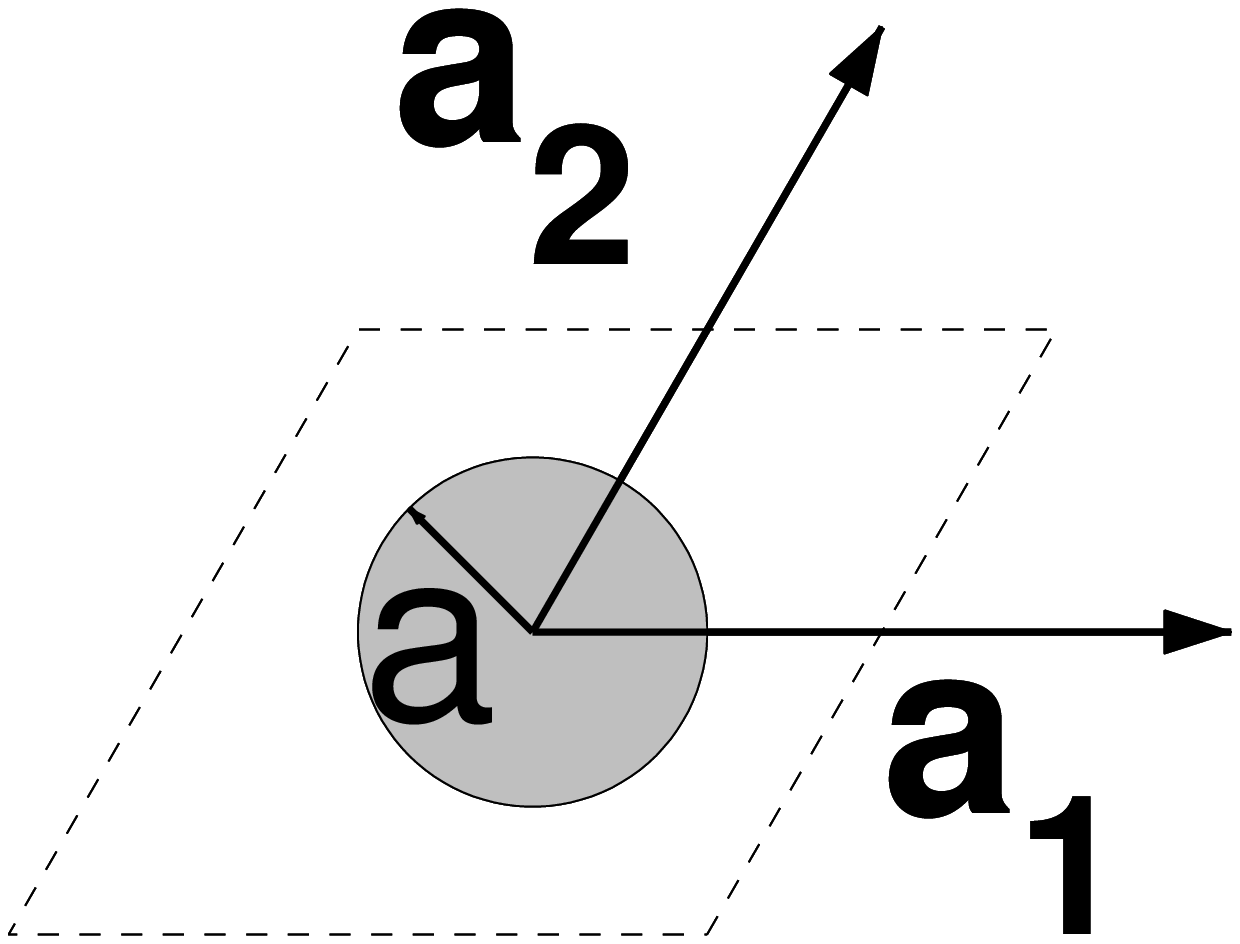}&$\frac{2\pi a^2}{\sqrt{3}d^2}$&$\frac{\pi}{2\sqrt{3}}$&$\frac{2\pi n_1}{d}$&$\frac{2\pi (2n_2-n_1)}{\sqrt{3}d}$\\
TH&\epsfxsize=0.95cm\epsffile{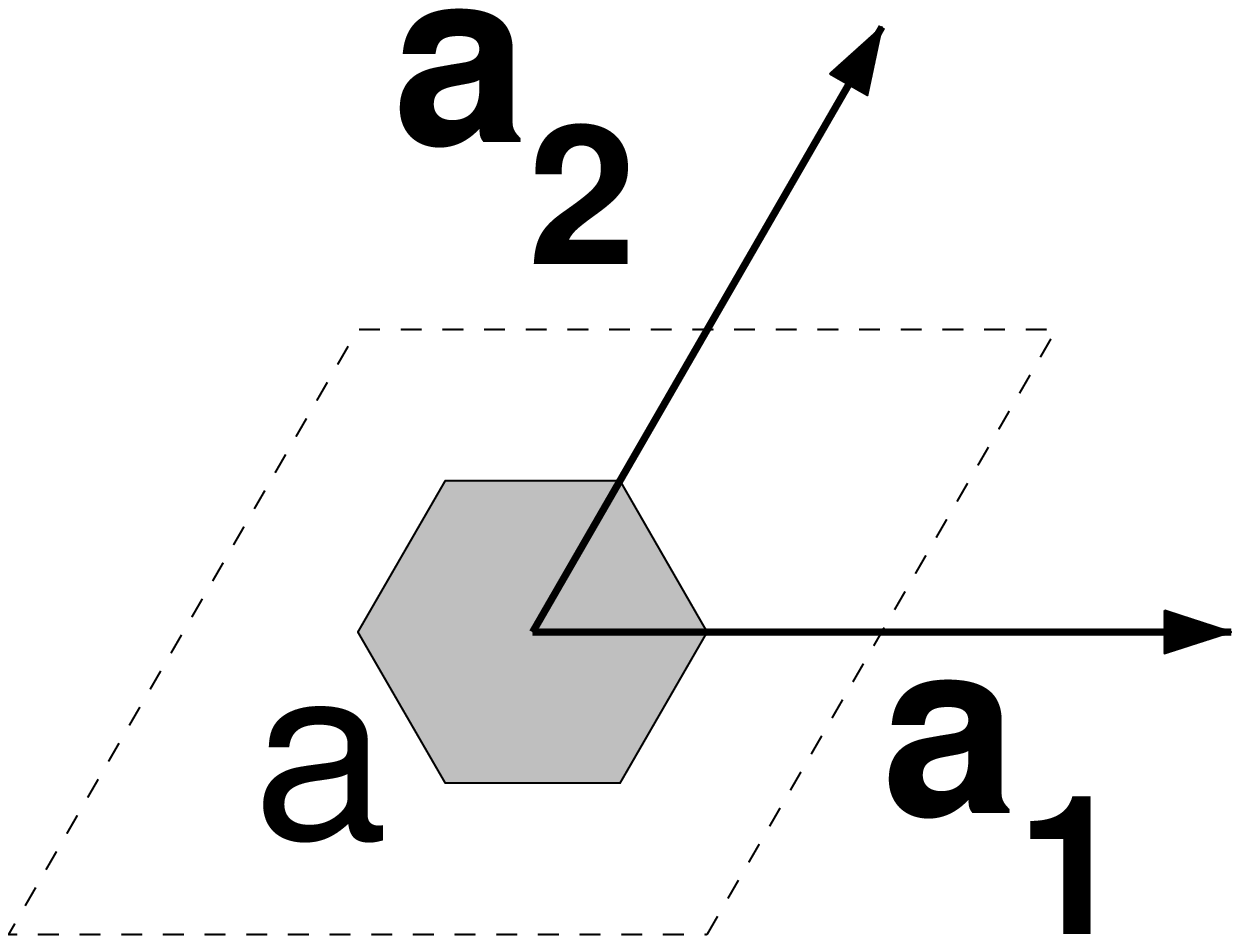}&$\frac{3a^2}{d^2}$&$\frac{3}{4}$&$\frac{2\pi n_1}{d}$&$\frac{2\pi (2n_2-n_1)}{\sqrt{3}d}$\\
TRH&\epsfxsize=0.95cm\epsffile{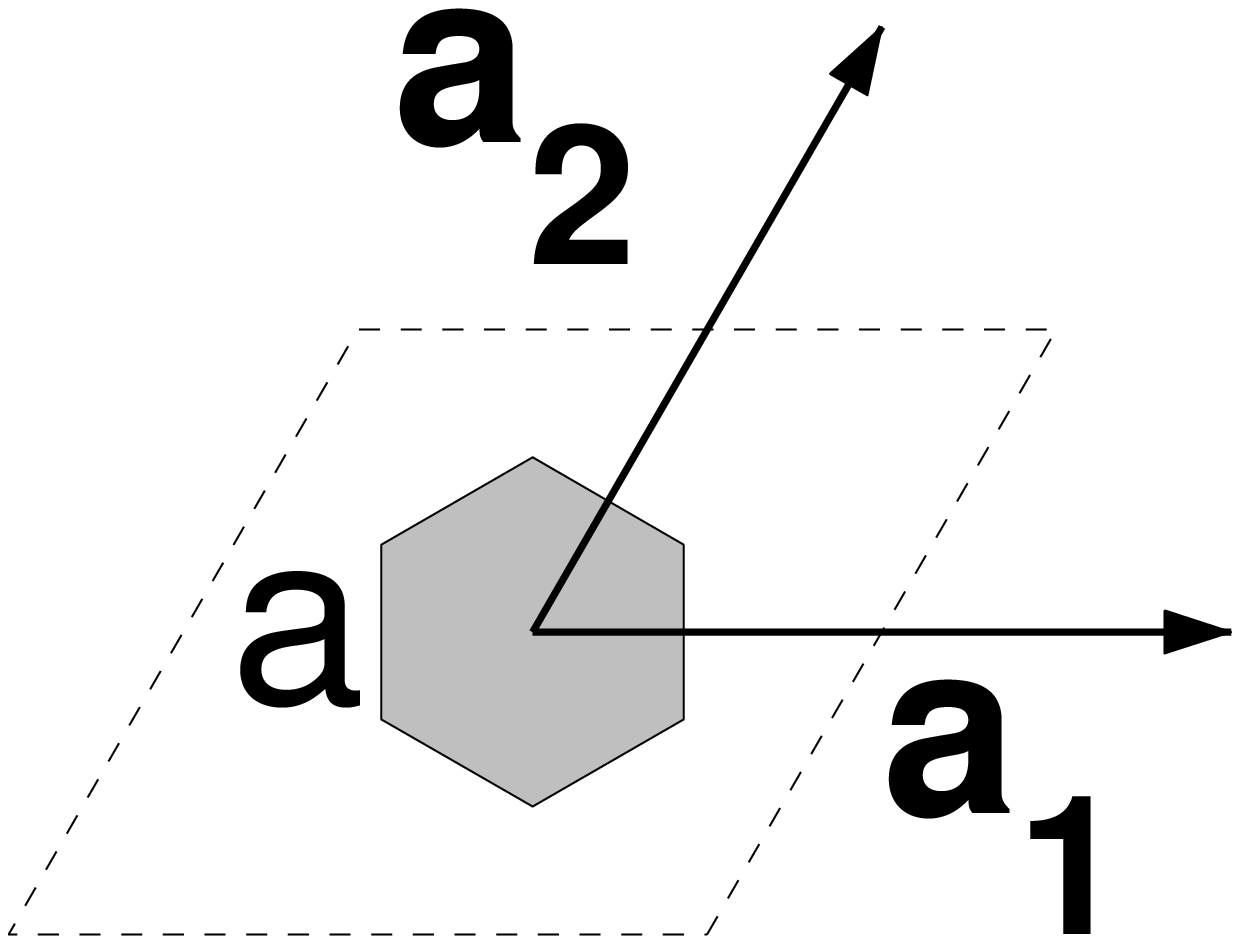}&$\frac{3a^2}{d^2}$&$1$&$\frac{2\pi n_1}{d}$&$\frac{2\pi (2n_2-n_1)}{\sqrt{3}d}$\\
TT&\epsfxsize=0.95cm\epsffile{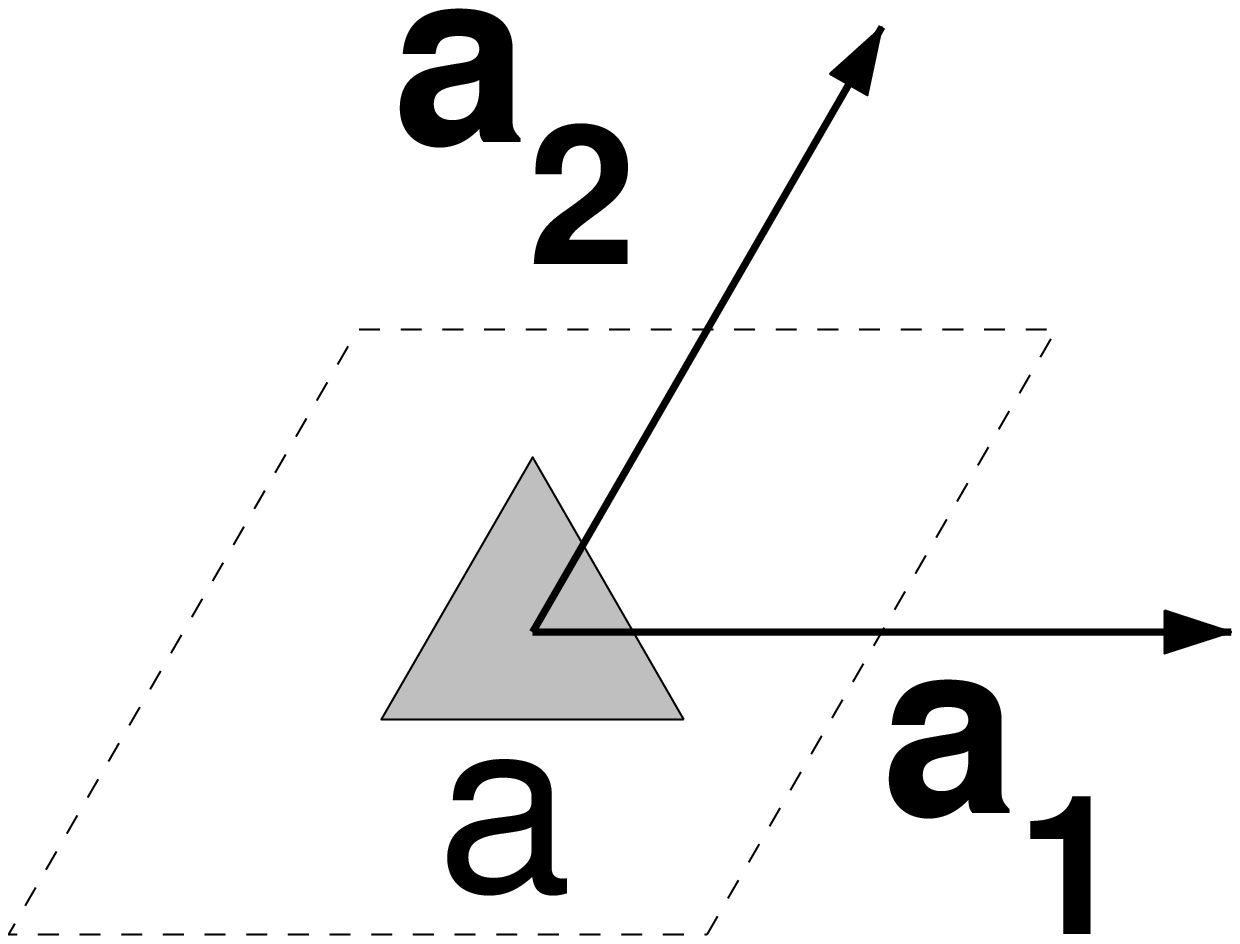}&$\frac{a^2}{2d^2}$&$\frac{1}{2}$&$\frac{2\pi n_1}{d}$&$\frac{2\pi (2n_2-n_1)}{\sqrt{3}d}$\\
HC&\epsfxsize=1.1cm\epsffile{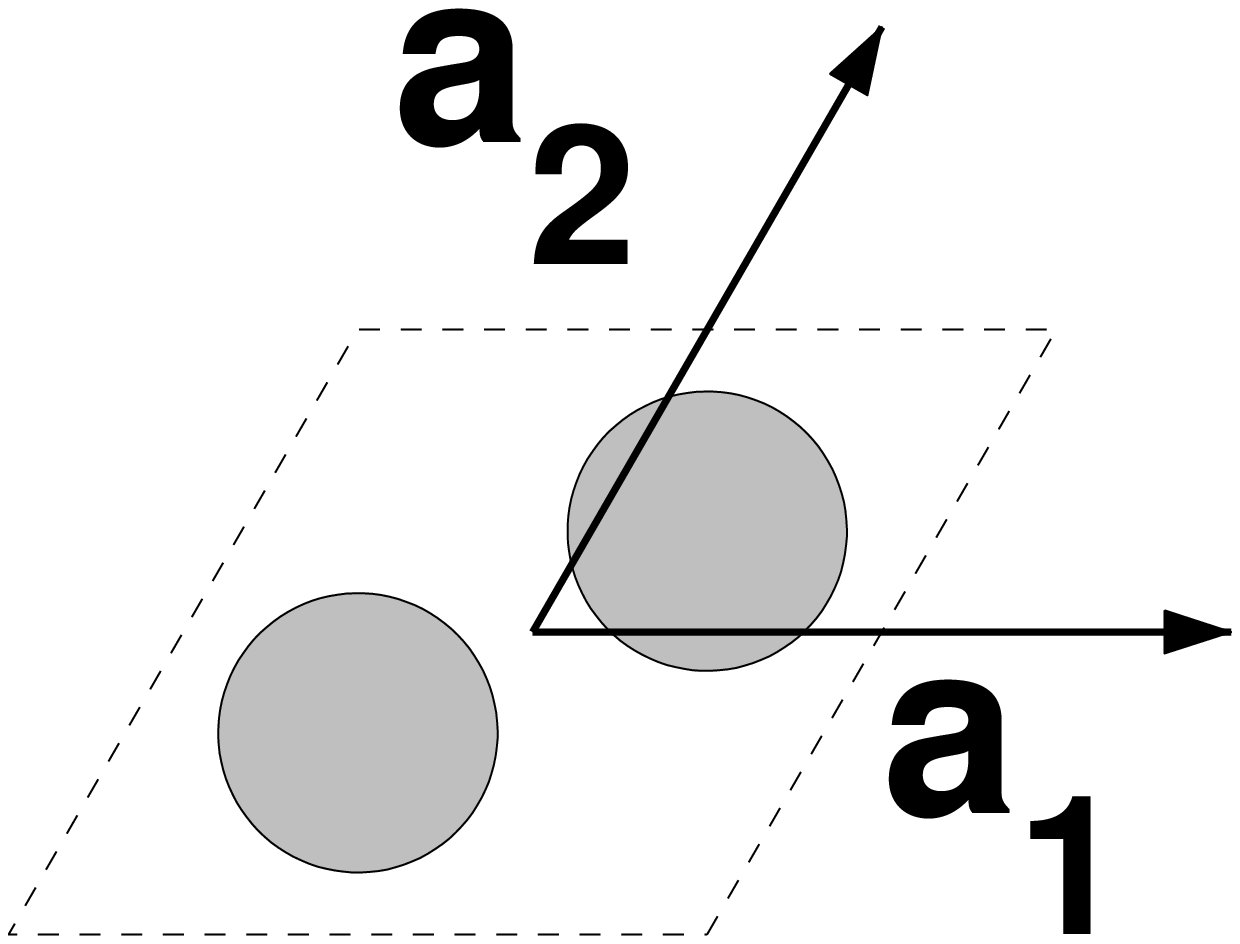}&$\frac{4\pi a^2}{\sqrt{3}d^2}$&$\frac{\pi}{3\sqrt{3}}$&$\frac{2\pi n_1}{d}$&$\frac{2\pi (2n_2-n_1)}{\sqrt{3}d}$\\
HH&\epsfxsize=1.1cm\epsffile{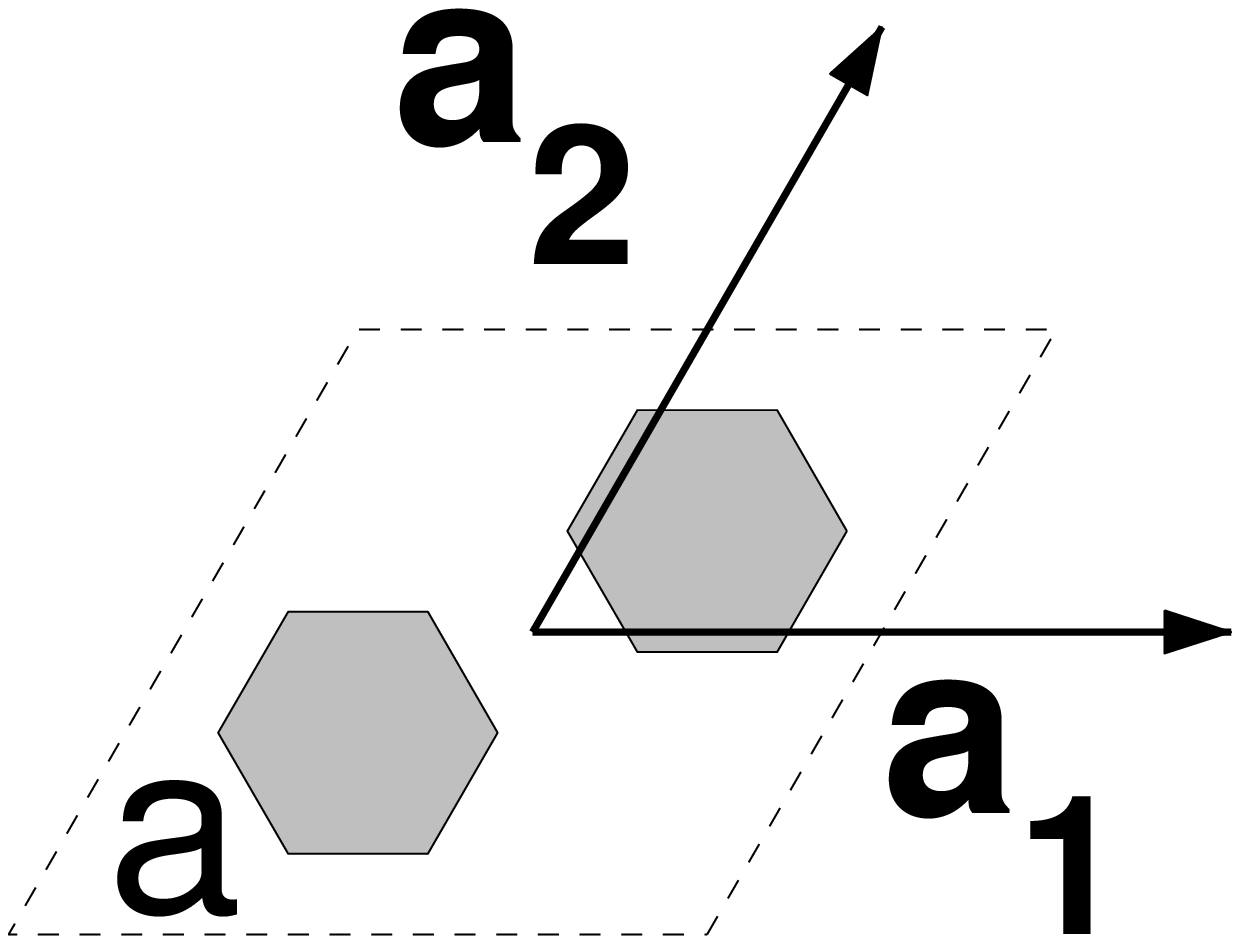}&$\frac{6a^2}{d^2}$&$\frac{2}{3}$&$\frac{2\pi n_1}{d}$&$\frac{2\pi (2n_2-n_1)}{\sqrt{3}d}$\\
HRH&\epsfxsize=1.1cm\epsffile{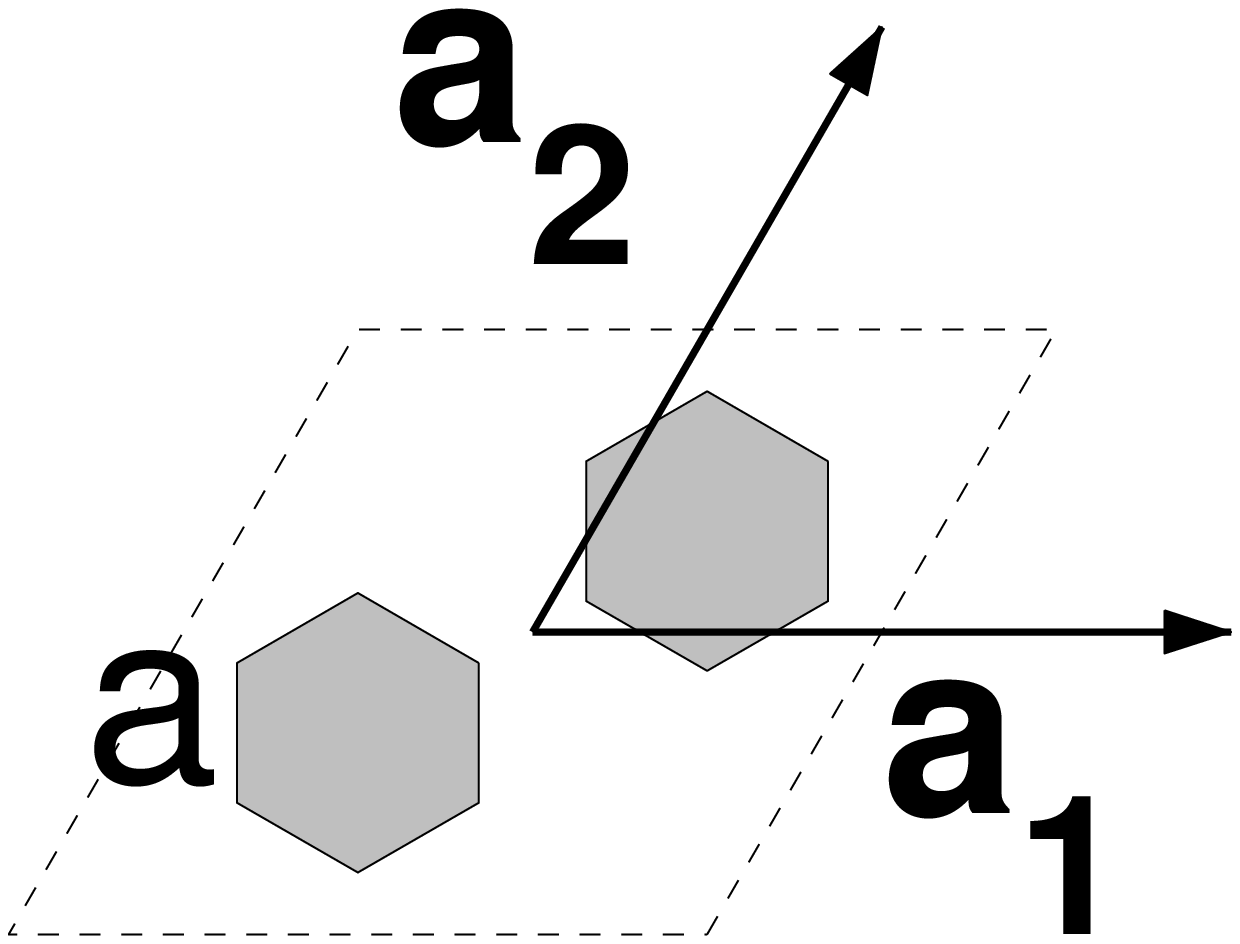}&$\frac{6a^2}{d^2}$&$\frac{1}{2}$&$\frac{2\pi n_1}{d}$&$\frac{2\pi (2n_2-n_1)}{\sqrt{3}d}$\\
HT&\epsfxsize=1.1cm\epsffile{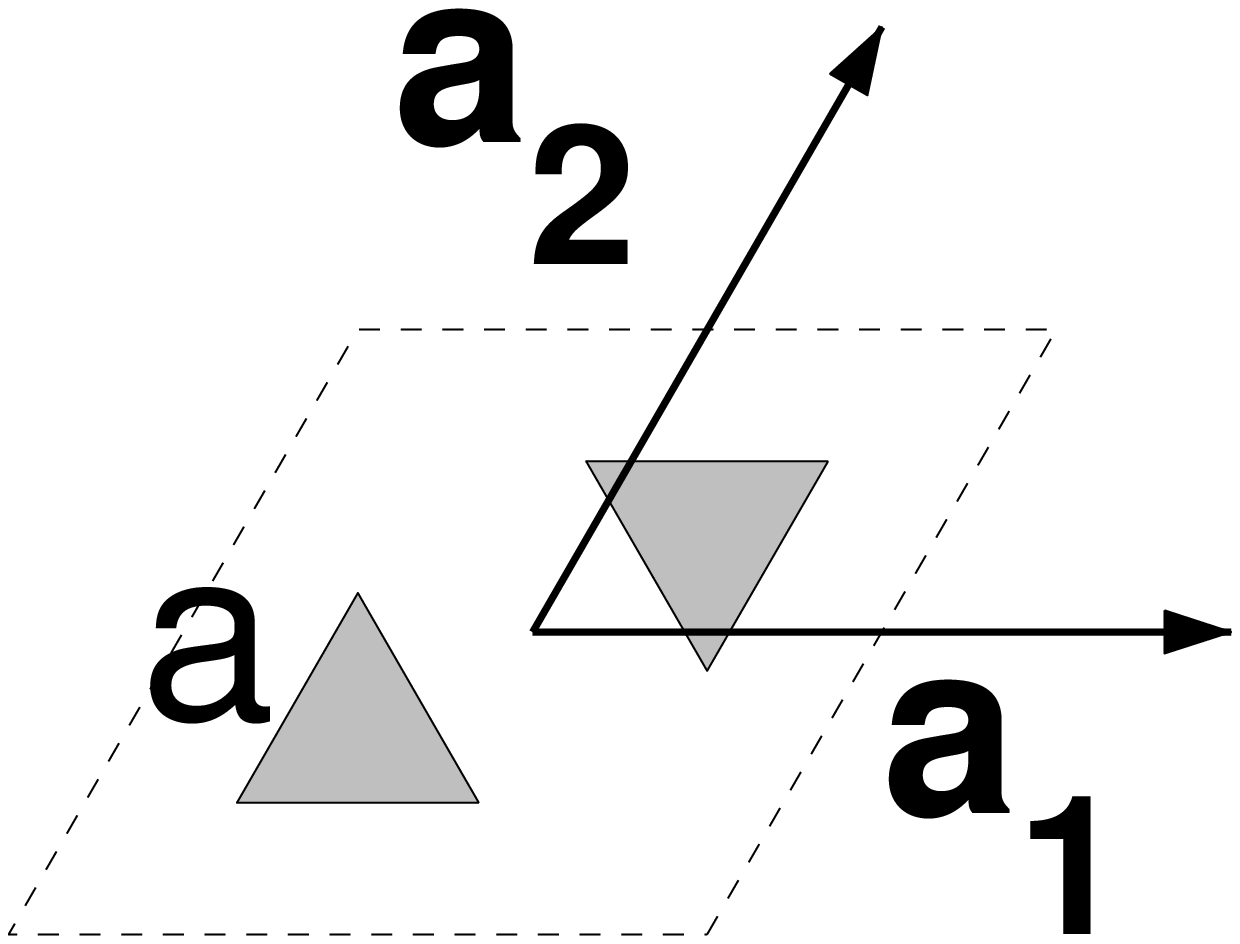}&$\frac{a^2}{d^2}$&$1$&$\frac{2\pi n_1}{d}$&$\frac{2\pi (2n_2-n_1)}{\sqrt{3}d}$\\
HRT&\epsfxsize=1.1cm\epsffile{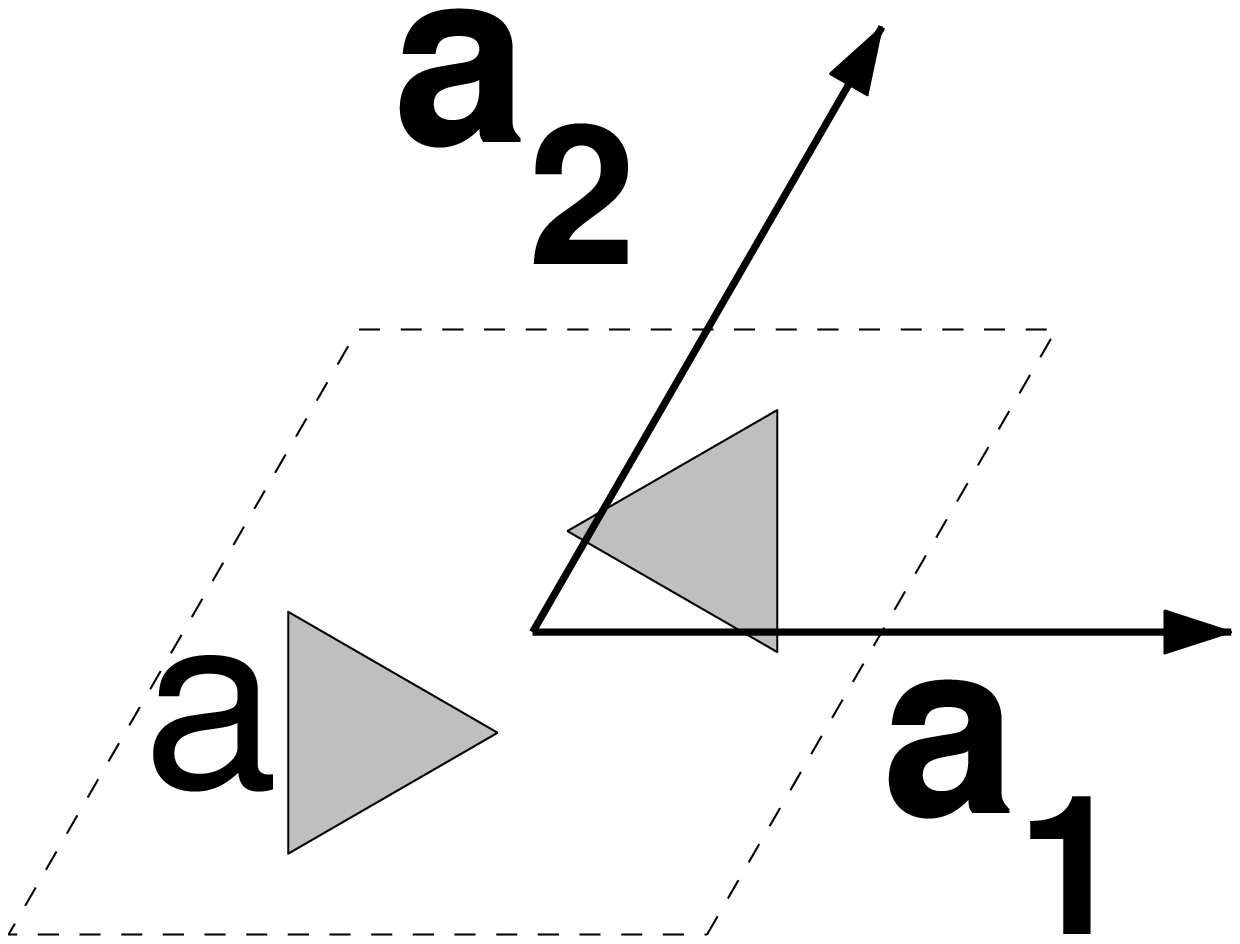}&$\frac{a^2}{d^2}$&$\frac{3}{4}$&$\frac{2\pi n_1}{d}$&$\frac{2\pi (2n_2-n_1)}{\sqrt{3}d}$\\
TD&\epsfxsize=1cm\epsffile{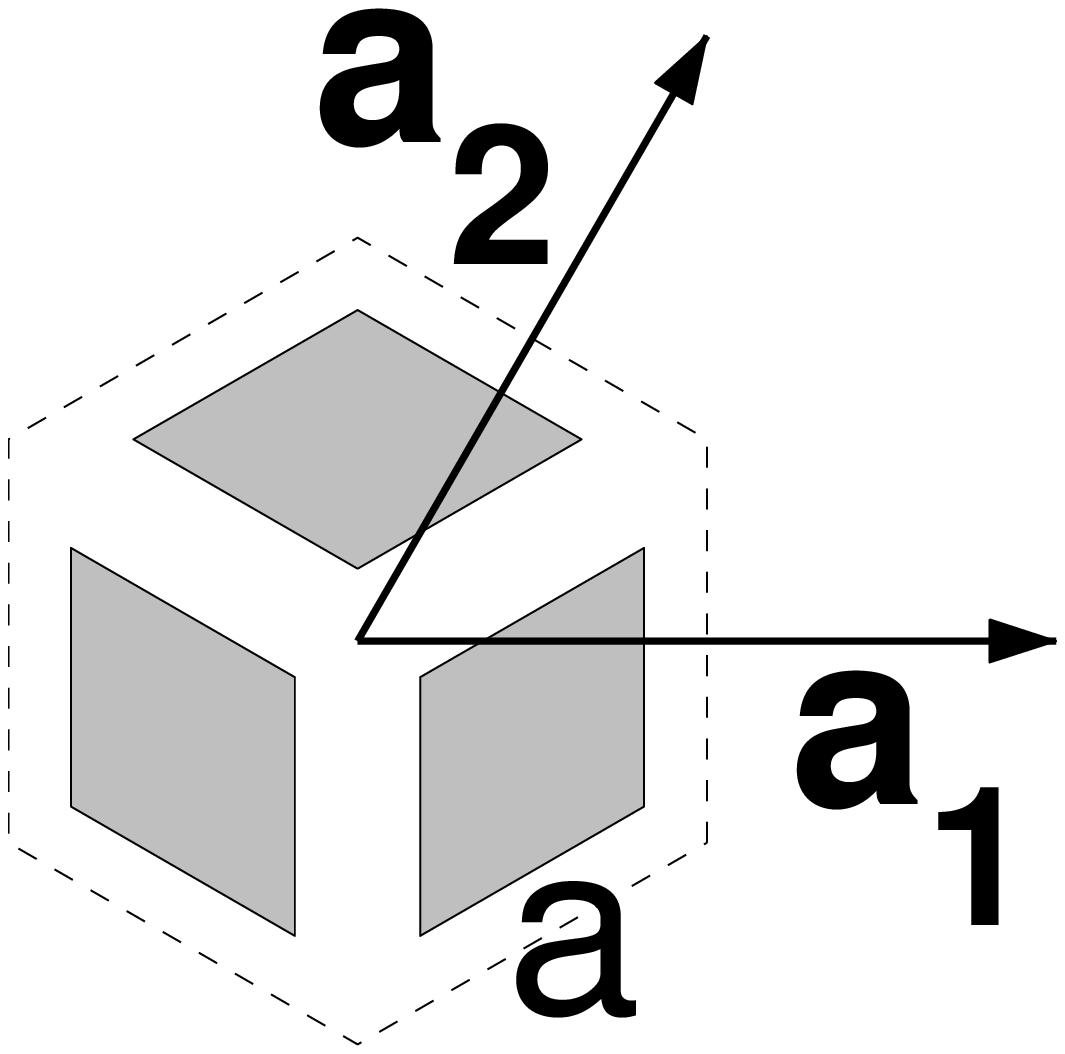}&$\frac{3a^2}{d^2}$&$1$&$\frac{2\pi n_1}{d}$&$\frac{2\pi (2n_2-n_1)}{\sqrt{3}d}$\\
TRD&\epsfxsize=1.4cm\epsffile{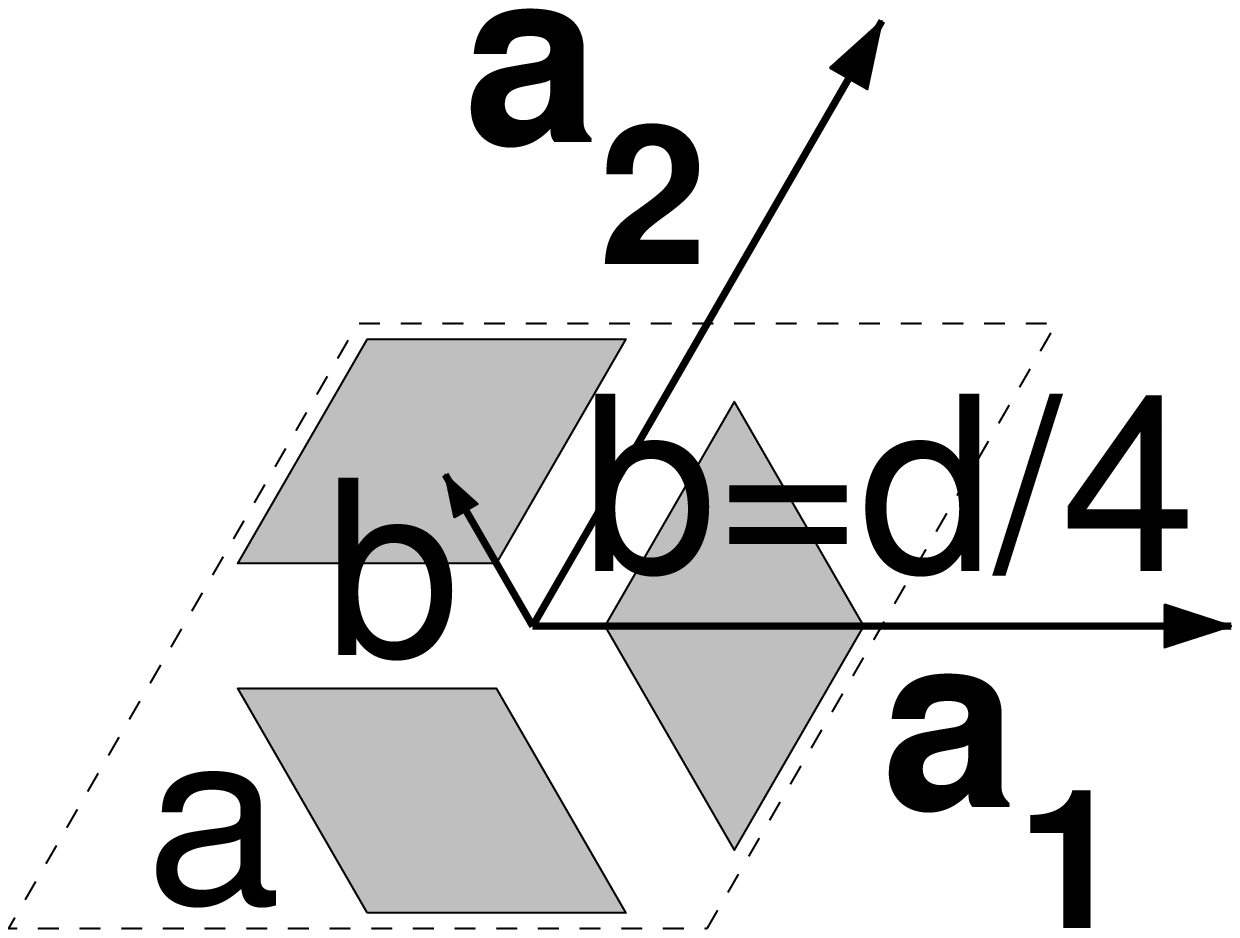}&$\frac{3a^2}{d^2}$&$\frac{3}{4}$&$\frac{2\pi n_1}{d}$&$\frac{2\pi (2n_2-n_1)}{\sqrt{3}d}$\\
&&&\\
\end{tabular}
\end{table}

\begin{table}
\label{tab2}
\caption{Structure Factors for Square and Triangular Lattices}
\begin{tabular}{ll}
{\Large$\Box$} latt. &$S({\mathbf G})$\\
\hline\\
SC&$2f\frac{J_1(Ga)}{Ga}$\\
SS&$fQ(\frac{G_xa}{2})
Q(\frac{G_ya}{2})$\\
SRS&$fQ
(\frac{(G_x+G_y)a}{2\sqrt{2}})Q
(\frac{(G_x-G_y)a}{2\sqrt{2}})$\\
SCR&$\frac{f}{2-\alpha}
\left[Q(\frac{G_xa}{2})Q(\frac{G_yb}{2})
+Q(\frac{G_xb}{2})Q(\frac{G_ya}{2})
-\alpha Q(\frac{G_xb}{2})Q(\frac{G_yb}{2})
\right]$\\
SRCR&$\frac{f}{2-\alpha}
\left[Q(\frac{(G_x+G_y)a}{2\sqrt{2}})Q(\frac{(G_x-G_y)b}{2\sqrt{2}})
+Q(\frac{(G_x+G_y)b}{2\sqrt{2}})Q(\frac{(G_x-G_y)a}{2\sqrt{2}})
-\alpha Q(\frac{(G_x+G_y)b}{2\sqrt{2}})Q(\frac{(G_x-G_y)b}{2\sqrt{2}})
\right]$\\
SMS&$\frac{f}{2}
\left[Q(\frac{G_xa}{2})Q(\frac{G_ya}{2})
+(-1)^{n_1+n_2}Q(\frac{(G_x+G_y)a}{2\sqrt{2}})
Q(\frac{(G_x-G_y)a}{2\sqrt{2}})
\right]$\\
SMCR&$\frac{f}{2(2-\alpha)}\left\{
\left[Q(\frac{G_xa}{2})Q(\frac{G_yb}{2})
+Q(\frac{G_xb}{2})Q(\frac{G_ya}{2})
-\alpha Q(\frac{G_xb}{2})Q(\frac{G_yb}{2})
\right]+(-1)^{n_1+n_2}
\left[Q(\frac{(G_x+G_y)a}{2\sqrt{2}})Q(\frac{(G_x-G_y)b}{2\sqrt{2}})
\right.\right.$\\
&$\left.\left.
+Q(\frac{(G_x+G_y)b}{2\sqrt{2}})Q(\frac{(G_x-G_y)a}{2\sqrt{2}})
-\alpha Q(\frac{(G_x+G_y)b}{2\sqrt{2}})Q(\frac{(G_x-G_y)b}{2\sqrt{2}})
\right]\right\}$\\&\\
\hline
{\large$\triangle$} latt.& $S({\mathbf G})$\\
\hline\\
TC&$2f\frac{J_1(Ga)}{Ga}$\\
TH&$\frac{2f}{3G_xa}
\left\{\sin\left[\frac{(3G_x-\sqrt{3}G_y)a}{4}\right]
Q\left[\frac{(G_x+\sqrt{3}G_y)a}{4}\right]
+\sin\left[\frac{(3G_x+\sqrt{3}G_y)a}{4}\right]Q\left[\frac{(G_x-\sqrt{3}G_y)a}{4}\right]\right\}$\\
TH($G_x=0$)&$\frac{2f}{3\left(\frac{\sqrt{3}G_ya}{2}\right)^2}
\left[1-\cos\left(\frac{\sqrt{3}G_ya}{2}\right)+\frac{\sqrt{3}G_ya}{2}
\sin\left(\frac{\sqrt{3}G_ya}{2}\right)\right]$\\
TRH&$\frac{2f}{3G_ya}
\left\{\sin\left[\frac{(\sqrt{3}G_x+3G_y)a}{4}\right]
Q\left[\frac{(\sqrt{3}G_x-G_y)a}{4}\right]
-\sin\left[\frac{(\sqrt{3}G_x-3G_y)a}{4}\right]Q\left[\frac{(\sqrt{3}G_x+G_y)a}{4}\right]\right\}$\\
TRH($G_y=0$)&$\frac{2f}{3\left(\frac{\sqrt{3}G_xa}{2}\right)^2}
\left[1-\cos\left(\frac{\sqrt{3}G_xa}{2}\right)+\frac{\sqrt{3}G_xa}{2}
\sin\left(\frac{\sqrt{3}G_xa}{2}\right)\right]$\\
TT&$\frac{2if\exp({\frac{-iG_ya}{4\sqrt{3}}})}{G_xa}
\left\{e^{\frac{-iG_xa}{4}}Q
\left[\frac{(G_x-\sqrt{3}G_y)a}{4}\right]
-e^{\frac{iG_xa}{4}}Q
\left[\frac{(G_x+\sqrt{3}G_y)a}{4}\right]\right\}$\\
TT($G_x=0$)&$\frac{2f\exp(\frac{iG_ya}{2\sqrt{3}})}
{\left(\frac{\sqrt{3}G_ya}{2}\right)^2}\left(1-i\frac{\sqrt{3}G_ya}{2}
-e^{-i\frac{\sqrt{3}G_ya}{2}}\right)$\\
HC&$2f\cos\left[\frac{(n_1+n_2)\pi}{3}\right]
\left(\frac{J_1(Ga)}{Ga}\right)$\\
HH&$\frac{2f\cos\left[\frac{(n_1+n_2)\pi}{3}\right]}{3G_xa}
\left\{\sin\left[\frac{(3G_x-\sqrt{3}G_y)a}{4}\right]
Q\left[\frac{(G_x+\sqrt{3}G_y)a}{4}\right]
+\sin\left[\frac{(3G_x+\sqrt{3}G_y)a}{4}\right]Q\left[\frac{(G_x-\sqrt{3}G_y)a}{4}\right]\right\}$\\
HH($G_x=0$)&$\frac{2f\cos\left[\frac{(n_1+n_2)\pi}{3}\right]}{3\left(\frac{\sqrt{3}G_ya}{2}\right)^2}
\left[1-\cos\left(\frac{\sqrt{3}G_ya}{2}\right)+\frac{\sqrt{3}G_ya}{2}
\sin\left(\frac{\sqrt{3}G_ya}{2}\right)\right]$\\
HRH&$\frac{2f\cos\left[\frac{(n_1+n_2)\pi}{3}\right]}{3G_ya}
\left\{\sin\left[\frac{(\sqrt{3}G_x+3G_y)a}{4}\right]
Q\left[\frac{(\sqrt{3}G_x-G_y)a}{4}\right]-\sin\left[\frac{(\sqrt{3}G_x-3G_y)a}{4}\right]Q\left[\frac{(\sqrt{3}G_x+G_y)a}{4}\right]\right\}$\\
HRH($G_y=0$)&$\frac{2f\cos\left[\frac{(n_1+n_2)\pi}{3}\right]}{3\left(\frac{\sqrt{3}G_xa}{2}\right)^2}
\left[1-\cos\left(\frac{\sqrt{3}G_xa}{2}\right)+\frac{\sqrt{3}G_xa}{2}
\sin\left(\frac{\sqrt{3}G_xa}{2}\right)\right]$\\
HT&$\frac{2f}{G_xa}
\left\{\sin\left[\frac{(\sqrt{3}G_x-G_y)a}{4\sqrt{3}}
+\frac{(n_1+n_2)\pi}{3}\right]
Q\left[\frac{(G_x+\sqrt{3}G_y)a}{4}\right]+
\sin\left[\frac{(\sqrt{3}G_x+G_y)a}{4\sqrt{3}}
-\frac{(n_1+n_2)\pi}{3}\right]
Q\left[\frac{(G_x-\sqrt{3}G_y)a}{4}\right]
\right\}$\\
HT($G_x=0$)&$\frac{2f}{\left(\frac{\sqrt{3}G_ya}{2}\right)^2}
\left\{\cos\left[\frac{G_ya}{2\sqrt{3}}
+\frac{(n_1+n_2)\pi}{3}\right]
+\frac{\sqrt{3}G_ya}{2}\sin\left[\frac{G_ya}{2\sqrt{3}}
+\frac{(n_1+n_2)\pi}{3}\right]-\cos\left[\frac{G_ya}{\sqrt{3}}
-\frac{(n_1+n_2)\pi}{3}\right]
\right\}$\\
HRT&$\frac{2f}{G_ya}
\left\{\sin\left[\frac{(G_x+\sqrt{3}G_y)a}{4\sqrt{3}}
-\frac{(n_1+n_2)\pi}{3}\right]
Q\left[\frac{(\sqrt{3}G_x-G_y)a}{4}\right]-\sin\left[\frac{(G_x-\sqrt{3}G_y)a}{4\sqrt{3}}
-\frac{(n_1+n_2)\pi}{3}\right]
Q\left[\frac{(\sqrt{3}G_x+G_y)a}{4}\right]
\right\}$\\
HRT($G_y=0$)&$\frac{2f}{\left(\frac{\sqrt{3}G_xa}{2}\right)^2}
\left\{\cos\left[\frac{G_xa}{2\sqrt{3}}
+\frac{(n_1+n_2)\pi}{3}\right]
-\cos\left[\frac{G_xa}{\sqrt{3}}
-\frac{(n_1+n_2)\pi}{3}\right]+\frac{\sqrt{3}G_xa}{2}\sin\left[\frac{G_xa}{2\sqrt{3}}
+\frac{(n_1+n_2)\pi}{3}\right]
\right\}$\\
TD&$\frac{f}{3}\left\{
e^{-i\frac{(2n_2-n_1)\pi}{3}}Q
\left[\frac{(\sqrt{3}G_x+G_y)a}{4}\right]
Q\left[\frac{(\sqrt{3}G_x-G_y)a}{4}\right]
+e^{i\frac{(n_1+n_2)\pi}{3}}
Q\left(\frac{G_ya}{2}\right)
Q\left[\frac{(\sqrt{3}G_x-G_y)a}{4}\right]\right.$\\
&
$\left.+e^{-i\frac{(2n_1-n_2)\pi}{3}}Q
\left(\frac{G_ya}{2}\right)
Q\left[\frac{(\sqrt{3}G_x+G_y)a}{4}\right]
\right\}$\\
TRD&$\frac{f}{3}\left\{i^{-n_1}Q
\left[\frac{(G_x+\sqrt{3}G_y)a}{4}\right]
Q\left[\frac{(G_x-\sqrt{3}G_y)a}{4}\right]+i^{(n_1-n_2)}Q
\left(\frac{G_xa}{2}\right)
Q\left[\frac{(G_x+\sqrt{3}G_y)a}{4}\right]+i^{n_2}Q
\left(\frac{G_xa}{2}\right)
Q\left[\frac{(G_x-\sqrt{3}G_y)a}{4}\right]
\right\}$\\ \\
\end{tabular}
\end{table}

\end{document}